\documentclass{ws-mpla}
\usepackage[super]{cite}
\usepackage{graphicx}
\usepackage{url}
\usepackage{subcaption}
\usepackage{hyperref}
\usepackage{lineno}

\newcommand{\eekt}{$ee$-$k_t$}

\begin{document}

\markboth{Hideki Okawa}{Quantum algorithms for pattern recognition at high-energy colliders}

\title{Quantum artificial intelligence for pattern recognition \\ at high-energy colliders: \\ 
Tales of Three ``Quantum's''}

\author{Hideki Okawa}

\address{Institute of High Energy Physics, Chinese Academy of Sciences, \\
Shijingshan, Beijing 100049,
China\\
okawa@ihep.ac.cn}

\maketitle

\begin{abstract}

    Quantum computing applications are an emerging field in high-energy  
    physics. Its ambitious fusion with artificial intelligence is expected to deliver 
    significant efficiency gains over existing methods and/or enable computation from 
    a fundamentally different perspective. High-energy physics is a big data science that utilizes
    large-scale facilities, detectors, high-performance computing, and its worldwide networks. 
    The experimental workflow consumes a significant amount of computing resources, and its annual
    cost will continue to grow exponentially at future colliders.
    In particular, pattern recognition is one of the most crucial and 
    computationally intensive tasks.  
    Three types of quantum computing technologies, i.e., quantum gates, quantum annealing, and 
    quantum-inspired, are all actively investigated for  
    high-energy physics applications, and each has its pros and cons. 
    This article reviews the current status of quantum computing applications for pattern 
    recognition at high-energy colliders.

\end{abstract}

\section{Introduction}	

The year 2025 marks the International Year of Quantum Science and Technology~\cite{qy2025}, 
celebrating the 100th anniversary of the initial development of quantum mechanics, 
in particular, Heisenberg's article ``Quantum-mechanical re-interpretation of kinematic and 
mechanical relations''~\cite{Heisenberg:1925zz}\footnote{This English translation of the 
original German title is taken from 
``Sources of Quantum Mechanics'', edited by B. L. van der Waerden (1967)}.
This work was established as matrix mechanics soon after, in collaboration with 
Born and Jordan~\cite{Born:1926uzf}. Schr\"{o}dinger expanded the matter-wave concept
introduced by De Broglie~\cite{deBroglie:1924ldk} and formulated 
quantum mechanics as wave mechanics~\cite{Schrodinger:1926xyk}. Dirac~\cite{Dirac:1927we}
and Jordan~\cite{Jordan:1927} independently proved that these two formulations are equivalent. 
It may not be a mere coincidence that the notion of spin was proposed in the same year, 1925,
by Uhlenbeck and Goudsmit~\cite{spin}. 

The first quantum revolution 
flourished in the 20th century through those paradigm shifts, enabling technologies 
such as lasers, transistors, magnetic resonance imaging, and semiconductors.
We are now in the midst of the second quantum revolution, in which, for the first 
time in human history, we can identify and control a single quantum bit. 
It led to the arrival of 
commercial quantum computers, and quantum supremacy was demonstrated 
by Google Sycamore~\cite{Arute:2019zxq} and Jiuzhang~\cite{Zhong:2020iql}
of the University of Science and Technology of China (USTC) in generating 
random states. 
However, we are still in the era of the so-called 
Noisy Intermediate-Scale Quantum (NISQ) computers~\cite{Preskill:2018jim}, and practical quantum 
advantage has yet to be established. Various distinctive technologies are being adopted for 
quantum computing hardware, and each has its pros and cons. 
A recent breakthrough by Google’s Willow chip~\cite{GoogleQuantumAIandCollaborators:2024efv} 
has changed the landscape towards error correction. However, its impact on future pathways 
to large-scale, error-tolerant quantum computers and their practical applications remains 
to be seen. The Zuchongzhi 3.0 quantum computer by USTC has achieved performance comparable 
to Willow~\cite{zuchongzhi3}. 

An urgent need and interest in applying quantum computing and algorithms to 
high-energy physics (HEP) arise from 
the fact that future colliders will face an enormous increase in datasets in the coming decades. 
For example, at the High Luminosity Large Hadron Collider (HL-LHC)~\cite{hl-lhc}, the computing time increases 
exponentially with pileup (additional proton-proton interactions per bunch crossing). It will 
reach exabyte-level data handling from the current petabyte level, leading to an annual cost 
increase by a factor of about 10 to 20~\cite{atlascomp2_art,cmscomp_art}. 
The $Z$-pole operation at the Circular Electron Positron 
Collider (CEPC)~\cite{CEPC_TDR,CEPC_RefTDR} proposed in China will face similar 
computational challenges. 
Quantum computing and algorithms may allow us to overcome some of those challenges. 

Applications of quantum computing to HEP can be categorized into two classes: 
(1) quantum machine learning and 
(2) quantum simulation\footnote{Be warned not to confuse with the simulation of 
quantum circuits by classical hardware. In this review, we will explicitly mention this case 
as a quantum circuit simulator.}. 
For the former, quantum properties such as the rich Hilbert space, superposition, 
entanglement (and tunneling for quantum annealing) could lead to improvement in learning data. 
For the latter, quantum computers simulate the time evolution of a given Hamiltonian; i.e., 
we can directly simulate quantum phenomena with quantum systems.

This article presents an overview of the brief history of quantum computing, the current
state of quantum computing technologies (both described in Section~\ref{sec:qc}), 
and its applications to experimental HEP, in particular for pattern recognition. 
As pattern recognition in HEP often amounts to optimization 
problems, Section~\ref{sec:Ising} describes the formulation of Ising and 
quadratic unconstrained binary optimization (QUBO) problems and how 
quantum algorithms can solve them. Then, the current status of quantum 
artificial intelligence applied to track reconstruction and jet 
clustering is summarized in 
Sections~\ref{sec:tracking} and~\ref{sec:jetreco}, respectively.  
Reviews covering wider ranges of applications in experimental and theoretical 
HEP can be found in 
Refs.~\refcite{Guan:2020bdl,Gray:2022fou,Delgado:2022hvo,Delgado:2022tpc,PRXQuantum.4.027001,qc4hep,Fang:2024ple}. 

\section{Quantum Computing}	
\label{sec:qc}

The concept of quantum computing emerged in the early 1980s from several physicists~\cite{qc_natrevphys}. 
In 1980, Manin proposed a quantum automaton based on superposition and 
entanglement~\cite{manin}, and Benioff put forward a microscopic quantum-mechanical 
Hamiltonian as a model of Turing machines~\cite{Benioff:1980fes}. 
In 1981, Feynman delivered a presentation on the possibility of simulating 
quantum-mechanical effects in quantum systems~\cite{Feynman:1981tf}, which is the 
conceptual emergence of quantum simulation.
Later on, Deutsch described a quantum Turing machine~\cite{deutsch} for the first time, 
paving the way for universal quantum computation. Development of innovative 
quantum algorithms has followed, such as Shor's factorization algorithm~\cite{Shor:1994jg},
which can crack the 
Rivest-Shamir-Adleman (RSA) encryption in polynomial time. 
Nakamura, Pashkin, and Tsai succeeded in realizing quantum bits in a superconducting 
material~\cite{Nakamura:1999mcw} for the first time. 
Kadowaki and Nishimori proposed quantum annealing~\cite{qa} in 1998, which led to 
the arrival of D-Wave's commercial quantum annealers in 2011. IBM introduced the 
first circuit-based commercial computers in 2019. 

\begin{figure}[t]
    \centering
        \includegraphics[width=\linewidth]{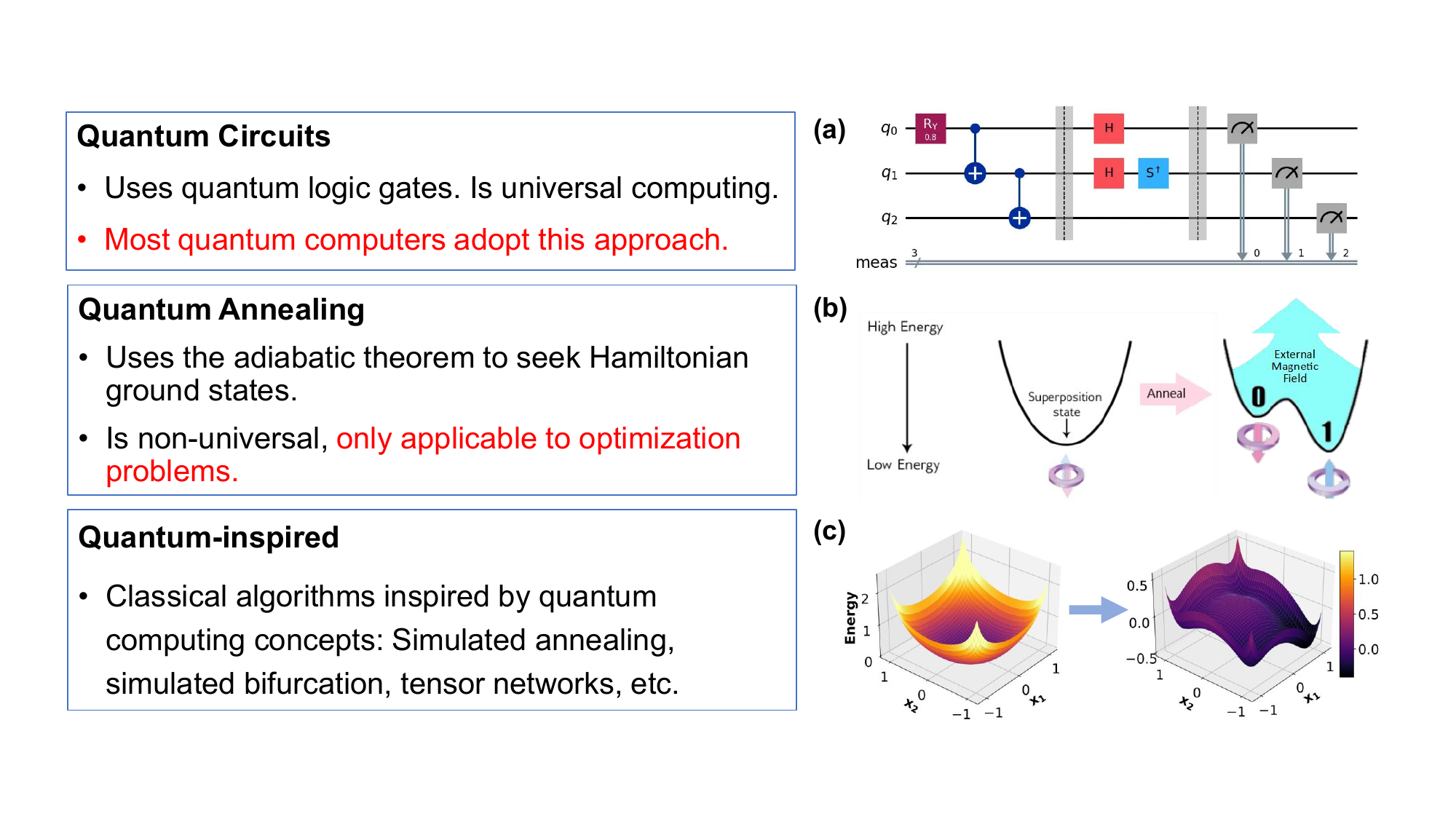}
        \caption{\label{fig:qctech} Three types of quantum computing technologies: 
        (a) quantum circuits, (b) quantum annealing 
        (From Ref.~\citenum{dwave-doc}, \href{https://creativecommons.org/licenses/by-nc-sa/4.0/}{CC BY-NC-SA 4.0}), 
        and (c) quantum-inspired (courtesy: Xianzhe Tao).}
\end{figure}

Quantum computing can be categorized into three major classes using distinctive techniques (Fig.~\ref{fig:qctech}): 
quantum circuits (gate-based), quantum annealing, and quantum-inspired. The former two use quantum hardware, 
whereas the last uses ``classical'' hardware but algorithms inspired by quantum-computational ideas. 
There is another class of models called analog quantum computing, which often involves 
continuous variables~\cite{Lloyd:1998jk}. 
However, its applications in HEP~\cite{Abel:2025zxb,Maier:2025ppr} are currently limited and 
have not been considered for pattern recognition. 

\subsection{Quantum circuits}

Quantum circuit hardware uses quantum logic gates, which perform unitary transformations.
It was first defined in Ref.~\refcite{qgate} and has been generalized in 
Ref.~\refcite{Nielsen_Chuang}. 
It exploits universal computing, and thus most quantum computers adopt this approach. 
Quantum gates can be realized using various technologies, including superconductors, ion traps, 
photons, neutral atoms, silicon chips, and diamond. The superconducting technique is the
most mature and scalable so far. Every approach has different performance and 
pros and cons in terms of the coherence longevity, processing speed, two-gate fidelity 
(i.e., logic success rate), connectivity, etc.   

\subsection{Quantum annealing}

Quantum annealing uses the adiabatic theorem to seek the ground state of a given 
Hamiltonian~\cite{qa}. It is non-universal computing and is only applicable to 
optimization problems described in Section~\ref{sec:Ising}. 
It is realized with a superconducting technique. Because of 
the fundamentally different implementation of the quantum circuits, quantum annealers
already employ thousands of qubits.  
In addition to the quantum properties used in quantum circuits, 
quantum tunneling helps avoid the ground state prediction from being trapped in local minima. 
Section~\ref{sec:qasolv}
describes the explicit formulation of how quantum annealing predicts the minimum energy 
of a Hamiltonian. 

\subsection{Quantum-inspired}

An important class of quantum-inspired approaches solves optimization 
problems via ``classical'' time evolutions of differential equations. Because of 
its resemblance to the quantum-annealing approach, it is also called quantum-annealing-inspired algorithms 
(QAIAs)~\cite{qaia}. Simulated annealing (SA)~\cite{SA}, simulated coherent Ising machines~\cite{SimCIM}, 
and simulated bifurcation (SB)~\cite{aSB,bSBdSB,HbSBHdSB} belong to this category. 
SA and SB have been considered in HEP applications; further 
details are provided in Section~\ref{sec:qaia}.

Another class of quantum-inspired algorithms actively investigated in high-energy
physics is tensor networks (TNs). They have been used to represent quantum many-body 
systems on classical computers, which have naturally become a key component for simulating 
quantum circuits on classical hardware. They have also been actively used in theoretical HEP
studies for particle dynamics. 
Their HEP applications to 
classification can be found in Refs.~\refcite{Borella:2024mgs,Abel:2025pxa,Araz:2021zwu}.
As QAIAs generally outperform TNs for optimization problems~\cite{qaia}, TNs have not been 
considered for pattern recognition in HEP and will not be covered in this review.

\section{Ising/QUBO Problems}	
\label{sec:Ising}

There is an important class of problems, the so-called combinatorial optimization problems. 
Practically complicated problems in society, such as route search, optimal distribution and 
placement, and scheduling, can be formulated as such problems.  
They can be mapped to Ising or QUBO
problems~\cite{Lucas}. The difference between the two lies in using the $\pm$1 spins for 
the former or zero/one binaries for the latter. 
The Ising/QUBO problems are non-deterministic polynomial time (NP) complete; i.e., 
no efficient algorithm exists to find the solution. 
However, Ising problem solvers can provide quasi-optimal answers in a reasonable amount of time. 
Quantum computers and algorithms could also play crucial roles in addressing these challenges. 

Solving the Ising problem is to find a spin configuration $\{x_i\}_{i=1}^N\in\{-1, 1\}^N$ that 
minimizes the Hamiltonian $H(x_i)$ of an Ising model with $N$ spins:
\begin{equation}
    H(x_i)=\frac{1}{2}\sum_{ij}^NJ_{ij}x_ix_j+\sum_i^Nh_ix_i, \label{eq:Ising}
\end{equation}
where $J_{ij}$ represents the spin-spin interactions and $h_i$ is the external 
field. Similarly, a QUBO Hamiltonian can be formulated as: 
\begin{equation}
	O_{\textrm{QUBO}}({s_i}) = \sum_{i,j=1}^{N} Q_{ij} s_i s_j,
	\label{eq:qubo}
\end{equation}
where $s_i$ is the binary \{0,1\} and $Q_{ij}$ is the QUBO matrix. 
The QUBO and Ising Hamiltonians can be converted 
to each other by:
\begin{align}
    x_i&\longleftrightarrow 2s_i-1,\\
    J_{ij}&\longleftrightarrow \frac{Q_{ij}}{2},\\
    h_i&\longleftrightarrow \frac{\sum_jQ_{ij}}{2}.
\end{align}
Various components of workflow in high-energy collider experiments can also be regarded as 
optimization procedures. 
Below are outlined algorithms for addressing optimization problems using the three quantum technologies.

\subsection{Solving Ising problems with quantum annealing}
\label{sec:qasolv}

Quantum annealing simulates the time evolution 
of the Transverse Field Ising Model relying on the adiabatic theorem (Fig.~\ref{fig:qctech} (b)), 
with the Hamiltonian $H$ given by:
\begin{equation}
    H(s) = A(s)H_\text{initial}+B(s)H_\text{target}, 
    \label{eq:Hqa}
\end{equation}
where $s$ is the normalized time parameter, and 
$H_\text{initial}$ is the initial Hamiltonian of the system, taken to be 
a transverse field, typically defined as:
\begin{equation}
H_\text{initial}=\sum_i\sigma_i^x.
\end{equation}
$\sigma_i^x$ is the Pauli $X$ matrix operating on the $i$-th qubit.
$H_\text{target}$ is the unknown ground state of the Ising Hamiltonian, 
representing the target optimization function: 
\begin{equation}
H_\text{target}=\sum_{ij}J_{ij}\sigma_i^z\sigma_j^z+\sum_ih_i\sigma_i^z,
\end{equation}
where $J_{ij}$ is the coupling strength between the $i$-th and $j$-th 
qubits, $\sigma_i^z$ is the Pauli $Z$ matrix operating on the 
$i$-th qubit, and $h_i$ is the strength of the external longitudinal
field applied to the $i$-th qubit.  
$A(s)$ and $B(s)$ are the annealing schedules of the Hamiltonian. 
Measuring the final state would provide the solution to the Ising problem. 

\subsection{Solving Ising problems with quantum circuits}

Quantum circuit machines can also solve Ising problems using variational circuits.
Quantum-classical hybrid methods such as Variational Quantum Eigensolver (VQE)~\cite{VQE}
and Quantum Approximate Optimization Algorithm (QAOA)~\cite{farhi2014quantum,farhi2015quantum} 
fall into this category. The principle of QAOA is described below as an example. 

QAOA mimics the quantum annealing procedure by performing Trotterization of the Hamiltonian
and searching for the ground state by scanning the variational parameters with classical 
optimizers. It consists of a quantum component that prepares a quantum state using 
a set of variational parameters, and a classical component that fine-tunes those 
parameters and returns them to the quantum component in a continuous feedback loop.  
The adiabatic evolution is defined the same as in Eq.~\ref{eq:Hqa}, with the linear annealing 
schedule ($A(s)=1-s$ and $B(s)=s$) usually adopted in QAOA. 
For each distinct parameter $p$, which corresponds to the circuit depth, 
sets of variational initial values for various QAOA circuit parameters are randomly generated. 
The Trotterization operation 
$U$ is pursued in sufficiently small time steps $\Delta t$: 
\begin{eqnarray}
U &\approx& \prod^{p}_{n=1} e^{-iH(n\Delta t)\Delta t} 
= \prod^{p}_{n=1} e^{-i\left(1-s(n\Delta t)\right)H_{\textrm{initial}}\Delta t} e^{-is(n\Delta t)H_{\textrm{target}}\Delta t} \nonumber \\
&=& \prod^{p}_{n=1} e^{-i \beta_n H_{\textrm{initial}}} e^{-i \gamma_n H_{\textrm{target}}} = \prod^{p}_{n=1} U_M(\beta_n) U_C(\gamma_n),
\end{eqnarray}
where $\beta_n$ and $\gamma_n$ are variational parameters, $U_C(\gamma_n)$ ($U_M(\beta_n)$) 
is the operation of consecutive cost (mixer) layers in QAOA. 
Executing the quantum circuits with the optimized variational parameters yields a 
specific quantum state that should correspond to the answer to the Ising problem. 
However, QAOA suffers from adiabatic bottlenecks and requires a large circuit. 
Scaling QAOA to large problems encounters Barren plateaus and 
is known to be challenging. 
A recent development of an imaginary Hamiltonian variational Ansatz (iHVA)~\cite{iHVA}
and Imaginary Time Evolution-Mimicking Circuit (ITEMC)~\cite{Chai:2025jnp}
overcomes those bottlenecks of QAOA, and they are validated up to a graph with 67 (80)
nodes for the former (latter). 

\subsection{Solving Ising problems with quantum-inspired algorithms}
\label{sec:qaia}
 
SA has been actively considered for optimization problems in HEP, 
recently as a benchmark to evaluate the effectiveness of quantum 
annealing, but also as a stand-alone option with dedicated hardware,
such as digital annealing~\cite{Saito:2020pld}.  Applications of SB in HEP have
recently emerged as a promising option~\cite{qaiatrack,qaiajet} due to its strong affinity 
with multiple processing and cutting-edge computing resources, such
as graphical processing units (GPUs) and field-programmable gate arrays (FPGAs). 
The mechanisms of these quantum-inspired algorithms
are briefly described below.

\subsubsection{Simulated annealing (SA)}

SA employs random movements throughout the solution space to search for the 
global minimum~\cite{SA}. 
Analogous to the process of determining the ground state of matter,
a ``temperature'' is introduced to 
initially ``melt'' a substance at a high temperature, and then, the temperature is gradually
decreased until the system ``freezes'', resembling the formation of a crystal from 
a molten state. 
A simple algorithm proposed in Ref.~\refcite{metropolis} is used for the iteration, 
which applies a small random displacement to the original spin configuration and 
computes the energy difference $\Delta E$.
If $\Delta E<0$, then the new configuration is immediately accepted as the 
temporary prediction, and the algorithm proceeds to the next iteration. 
Conversely, if $\Delta E > 0$, the new configuration is accepted only probabilistically, 
according to the Boltzmann factor $P(\Delta E)=\exp(-\Delta E/k_B T)$, which is 
introduced to avoid local minima.  
The process continues until the energy stabilizes or is terminated by a user-defined
duration. 

\subsubsection{Simulated bifurcation (SB)} 

SB addresses combinatorial optimization problems via the 
adiabatic evolution of Kerr-nonlinear parametric oscillators (KPOs), 
which exhibit bifurcations corresponding to the two Ising spin states. 
The real and imaginary parts of the KPO amplitudes are interpreted as the 
positions $x_i$ and momenta $y_i$ of virtual particles. Taking the sign of 
the particle positions, sgn$(x_i)$, determines the spin.   
SB can update all oscillators in parallel, thereby accelerating computation. 
This feature is in clear contrast to SA, 
which generally cannot update spins independently. 
However, the original version of SB, adiabatic SB (aSB)~\cite{aSB}, 
is prone to errors 
originating from the continuous treatment of the spins $x_i$  
in the differential 
equations. To mitigate these analog errors, two variants of SB are introduced: 
ballistic SB (bSB) and discrete SB (dSB)~\cite{bSBdSB}. 
Inelastic walls are introduced at $x_i=\pm1$ for both, and 
$x_j$ is discretized to sgn$(x_j)$ in the mean-field term for dSB 
to suppress the error from the continuous relaxation of $x_i$:
\begin{eqnarray}
    \centering
\dot{x}_i &=& \frac{\partial H_{\textrm{SB}}}{\partial y_i} = a_0 y_i,\\
\dot{y}_i &=& - \frac{\partial H_{\textrm{SB}}}{\partial x_i}, \nonumber \\
      &=& \begin{cases}
        - \left[ a_0 - a(t) \right] x_i + c_0 \left( h_i + \sum_{j=1}^{N} J_{ij}x_j \right), \text{[bSB]}  \\
        -[a_0-a(t)]x_i+c_0\left(\sum_{j=1}^NJ_{ij}\text{sgn}(x_j)+h_i\right), \text{[dSB]}
          \end{cases} \label{eq:SB}  \\
\end{eqnarray}
where $a_0$ is the positive detuning frequency, $c_0$ represents the positive coupling strength, 
and $a(t)$ denotes a time-dependent pumping amplitude that monotonically increases
from zero to $a_0$. 
$J_{ij}$ and $h_i$ are the spin-spin interactions and the external field, 
respectively, given by the Ising problem (Eq.~\ref{eq:Ising}) under consideration.
The Hamiltonian of the non-linear oscillator
system $H_{\text{SB}}$ and its potential energy $V_{\text{SB}}$ are defined as: 
\begin{eqnarray}
H_{\textrm{SB}} &=& \frac{a_0}{2} \sum_{i=1}^{N} y_i^2 + V_{\textrm{SB}},  \\
V_{\textrm{SB}} &=& \begin{cases}
            \frac{a_0-a(t)}{2}\sum_{i=1}^Nx_i^2-c_0\left(\frac{1}{2}\sum_{i,j}^NJ_{ij}x_ix_j+\sum_ih_ix_i\right), \text{[bSB]}\\
            \frac{a_0-a(t)}{2}\sum_{i=1}^Nx_i^2-c_0\left(\frac{1}{2}\sum_{i,j}^NJ_{ij}x_i\text{sgn}(x_j)+\sum_j^Nh_ix_i\right), \text{[dSB]} \\
            \quad \forall x_i, \ |x_i|\leq 1 \\
            \infty, \ \text{otherwise}  \quad \text{[bSB, dSB]}
            \end{cases}
\end{eqnarray}
Finally, as stated above, sgn($x_i$) gives the solution to the Ising problem. 
For all SB variants, numerical computations are performed using the 
symplectic Euler method~\cite{Leimkuhler_Reich}. 
Another set of variants that introduces heat fluctuations~\cite{HbSBHdSB} has 
a relatively mild impact on the performance and is currently not 
considered in HEP. 

\section{Track reconstruction}
\label{sec:tracking}

\begin{figure}[t]
    \centering
      \begin{subfigure}{0.5\linewidth}
            \includegraphics[width=\linewidth]{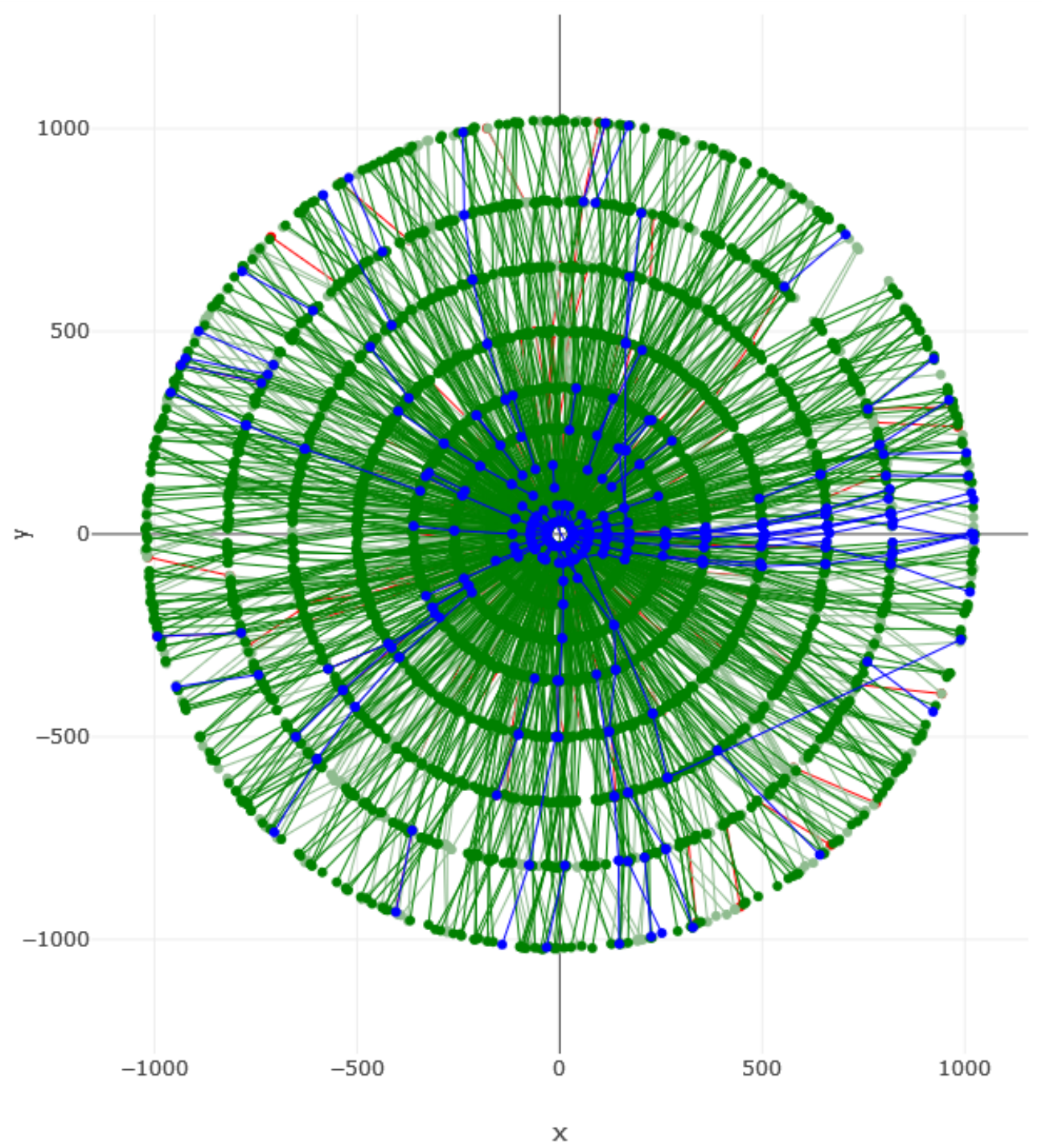}
      \end{subfigure}
      \caption{\label{fig:trk} Display of reconstructed tracks in an event from the 
      TrackML dataset~\cite{Amrouche:2019wmx,Amrouche:2021nbs}, which is simulated
      under the HL-LHC conditions. The green (red) lines represent tracks that are correctly 
      (incorrectly) reconstructed, whereas the blue lines are unreconstructed tracks.
      Reproduced from Ref.~\citenum{qaiatrack}. \href{https://creativecommons.org/licenses/by/4.0/}{CC BY 4.0}.} 
\end{figure}

Track reconstruction or tracking is a pattern recognition procedure to 
recover trajectories of charged particles and measure their momenta 
from hits in the silicon detector layers and/or drift chambers (Fig.~\ref{fig:trk}). 
Track reconstruction algorithms determine how to assign the detector hits to form 
track candidates, i.e. so-called track finding, and then pursue track fitting 
to obtain helix parameters. 
The combined Kalman filter is actively used in various experiments for track
finding and is adopted in A Common Tracking Software (ACTS)~\cite{acts}, 
for example. Recently, machine learning approaches such as Graph Neural Networks 
(GNNs)~\cite{ExaTrkX:2021abe,Jia:2024rbx,Reuter:2024kja}
and Transformers~\cite{Caron:2024cyo,Caron:2025yaa,md46-yqgd,trt} are also gaining 
attention for track finding. 
The least squares ($\chi^2$) or Runge-Kutta method is frequently 
used for track fitting. Those tracks are then used to reconstruct 
vertices and physics analysis objects further. 

Applications of quantum algorithms can be categorized into two classes of 
reconstruction: 
(1) a global method that formulates track finding as an Ising/QUBO problem and 
(2) an iterative method that attempts to replace existing classical 
computation by quantum algorithms. 

\subsection{Global track finding as an Ising problem}
\label{sec:globalTrk_Ising}

Formulating track finding as an optimization problem dates back to the Denby-Peterson 
method~\cite{Denby:1987rk,Peterson:1988gs}, which used neural networks or cellular automata 
to predict the ground state of a proposed Hamiltonian resembling a QUBO. 
The method was later employed at the ALEPH~\cite{oldqubotracking}, ARES~\cite{Baginian:1994gb}, 
ALICE~\cite{Pulvirenti:2004fq}, and LHCb~\cite{Passaleva:2008yda} experiments.  
In those studies, all potential combinations of two detector hits, i.e., doublets, were generated
and subjected to some quality criteria. The algorithm performed binary classification  
to determine which doublet combinations should be retained to form track candidates.
Two pioneering works applying quantum algorithms used 
quantum annealing~\cite{Zlokapa:2019tkn,Bapst:2019llh} with either doublets 
or triplets (i.e., segments of three detector hits) to reconstruct tracks. 

Ref.~\refcite{Zlokapa:2019tkn} adopted a doublet-based approach and modified the 
Denby-Peterson Hamiltonian~\cite{Denby:1987rk,Peterson:1988gs} to match the 
HL-LHC configuration as:
\begin{align}
\label{eq:DPqubo}
\begin{split}
E = &-\sum_{a, b, c} \left(\frac{\cos^\lambda(\theta_{abc})+\rho \cos^\lambda(\phi_{abc})}{r_{ab} + r_{bc}}\right)s_{ab}s_{bc}
+\eta\sum_{a, b c} \left(z_c - \frac{z_c - z_a}{r_c - r_a}r_c\right)^\zeta s_{ab}s_{bc}\\
&+\alpha\left(\sum_{b\neq c} s_{ab} s_{ac} + \sum_{a\neq c} s_{ab} s_{cb} \right) 
-\sum_{a,b}\left(\beta P(s_{ab}) - \gamma\right)s_{ab},
\end{split}
\end{align}
where $E$ is the QUBO Hamiltonian; $a$, $b$, $c$ are the detector-hit labels 
for the two connected doublets; 
$\theta_{abc}~(\phi_{abc})$ is the Cartesian (azimuthal) angle between 
the doublets; $r_{ab}$, $r_{bc}$ are the doublet lengths; $s_{ab}$, $s_{bc}$ are the 
binary variables to define whether those doublets should be kept; $z$, $r$ are the hit 
location in the $rz$-plane\footnote{The $z$-axis is chosen to align with the beam 
direction at the collider experiments, and $r$ is the distance in polar coordinates 
in the plane perpendicular to the beam axis.}; $P(s_{ab})$ is the prior probability for the given 
doublet using a Gaussian kernel density estimation (KDE); and $\lambda$, $\rho$, 
$\eta$, $\zeta$, $\alpha$, $\beta$, $\gamma$ are parameters defined in Table~\ref{tab:params}, 
determined by Bayesian optimization. 
\begin{table}[t]
\centering
\caption{Parameters that enter the definition of the modified Denby-Peterson QUBO (Eq.~\ref{eq:DPqubo})~\cite{Zlokapa:2019tkn}.
The description corresponds to what the parameters are designed for.
}\label{tab:params}
\begin{tabular}{ c|c|c }
 Parameter & Value & Description\\ 
 \hline
 $\lambda$ & 13.17 & Track angle separator\\
 $\rho$ & 5.00 & High-momentum bias\\
 $\eta$ & 14.41 & Beam spot bias\\
 $\zeta$ & 1.79 & Beam spot separator\\
 $\alpha$ & 86.20 & Bifurcation penalty\\
 $\beta$ & 20.91 & Edge alignment penalty\\
 $\gamma$ & 9.79 & Total edge count penalty
\end{tabular}
\end{table}
The parameters are chosen to provide the 
best performance with SA in terms of the harmonic mean of track reconstruction 
efficiency and purity, defined as: 
\begin{equation}
	\mathrm{Efficiency~(Recall)} = \frac{\textrm{The number of correctly reconstructed doublets}}{\textrm{The number of true doublets}}, \label{eq:efficiency}
\end{equation}
\begin{equation}
\centering
	\mathrm{Purity~(Precision)} = \frac{\textrm{The number of correctly reconstructed doublets}}{\textrm{The number of all reconstructed doublets}}. \label{eq:purity}
\end{equation}
The track reconstruction performance was evaluated using the TrackML Challenge
dataset~\cite{Amrouche:2019wmx,Amrouche:2021nbs}. This dataset simulates 
a top-quark pair process generated with Pythia8~\cite{pythia8} with additional 200 
proton-proton interactions overlaid that align with the HL-LHC conditions,
along with a fast detector simulation. 
The D-Wave 2X quantum annealer with 1,098 qubits and SA demonstrated promising 
and comparable track reconstruction efficiency and purity up to 500 particles.
Track finding for events with more particles could not be pursued due to the 
limited number of qubits and connectivity in D-Wave hardware. The SA 
results up to 1,000 particles showed that the track reconstruction performance
does not degrade with higher multiplicity. 

LHCb recently investigated a similar doublet-based method using a Denby-Peterson Hamiltonian~\cite{Nicotra:2023rmn}.
They first evaluated the method with a classical solver on $B_s \rightarrow \phi\phi$ events 
using the full detector simulation and obtained a performance comparable to 
the state-of-the-art approach at the LHCb, 
i.e., Search-by-Triplet~\cite{SbT_LHCb}. 
Then, they moved on to testing the quantum Harrow-Hassadim-Lloyd (HHL) algorithm~\cite{Harrow:2008dvn} 
for the Hamiltonian minimization for events up to five particle multiplicity  
with a simple detector toy model~\cite{codeZenodo}. 
A noiseless quantum circuit simulator, Qiskit Aer, pursued the HHL algorithm 
up to 14 qubits, and the quantum circuit was designed to fit 
the \texttt{ibm\_hanoi} 27-qubit r5.11 Falcon quantum processor. 
Hit purity and efficiency of 100\% can be achieved for the simple toy model events, 
but require a few thousand to tens of millions of circuit depth, 
which is unfeasible for current circuit-based quantum hardware.  

\begin{figure}[t]
    \centering
      \begin{subfigure}{0.9\linewidth}
          \includegraphics[width=\linewidth]{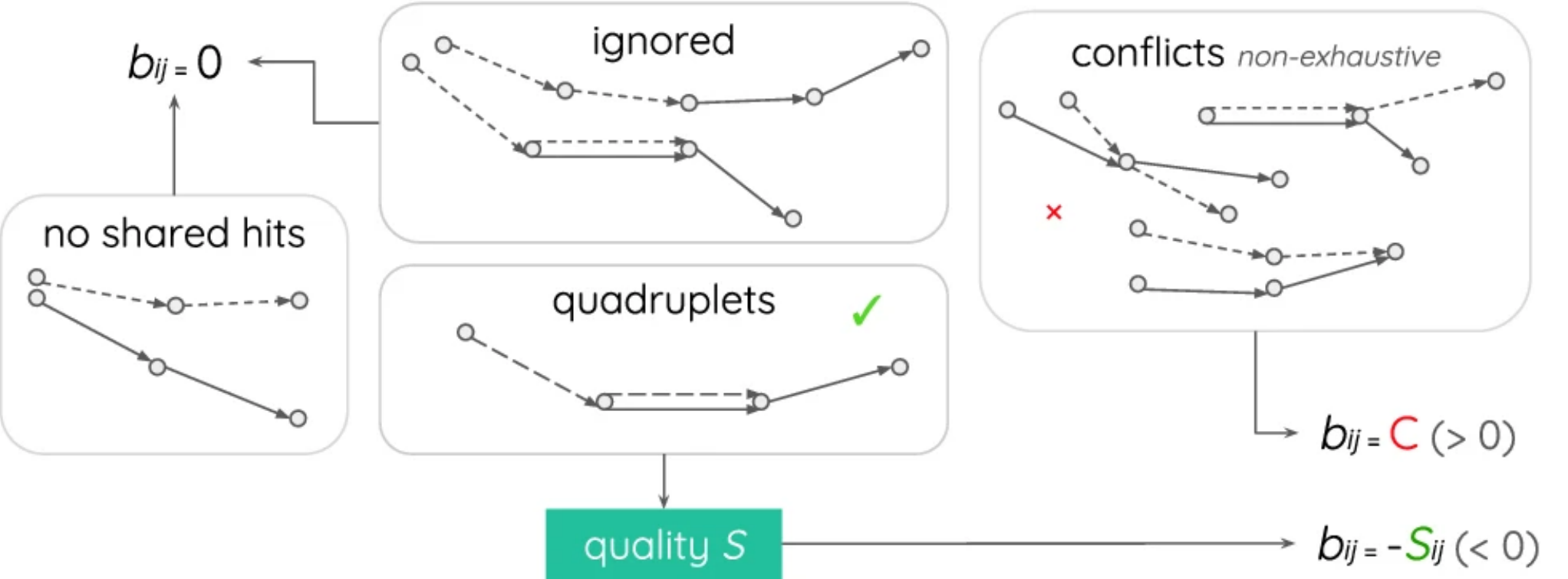}
      \end{subfigure}
      \caption{\label{fig:quboTrk} Criteria for assigning QUBO quadratic coefficients depending on the 
      configurations of triplet pairs. Reproduced from Ref.~\citenum{Bapst:2019llh}. \href{https://creativecommons.org/licenses/by/4.0/}{CC BY 4.0}.} 
  \end{figure}

\begin{figure}[t]
    \centering
      \begin{subfigure}{0.9\linewidth}
            \includegraphics[width=\linewidth]{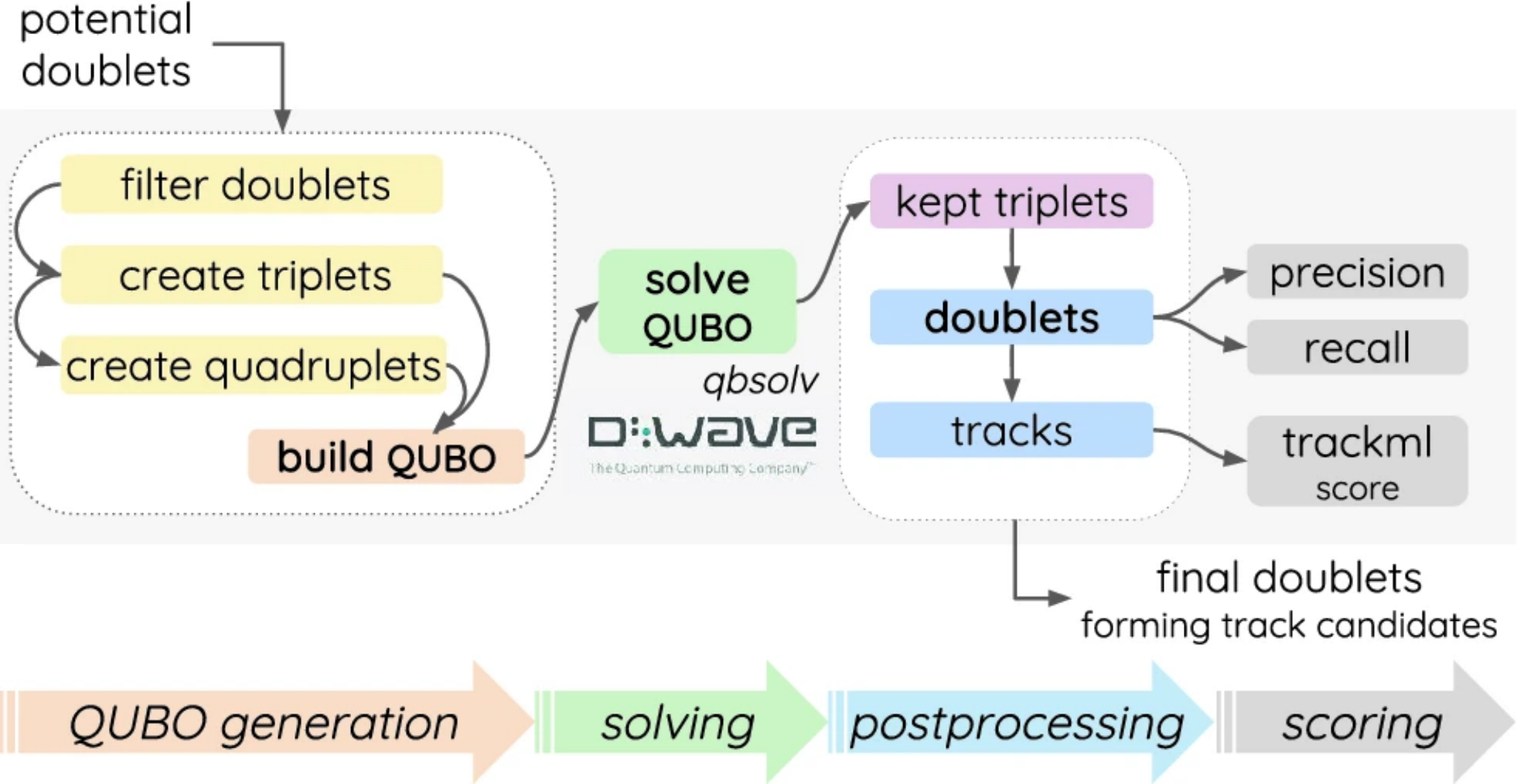}
      \end{subfigure}
      \caption{\label{fig:trk_workflow} Schematic overview of tracking workflow using the QUBO
      formulation (b). Reproduced from Ref.~\citenum{Bapst:2019llh}. \href{https://creativecommons.org/licenses/by/4.0/}{CC BY 4.0}.} 
  \end{figure}

The other pioneering quantum annealing approach adopted 
triplets instead~\cite{Bapst:2019llh}. The triplet-based QUBO Hamiltonian is defined as: 
\begin{equation}
        O(a,b,T) = \sum_{i=1}^{N} a_i T_i + \sum_{i=1}^{N} \sum_{j<i}^{N} b_{ij} T_i T_j,
        \label{eq:qubo_triplet}
\end{equation}
where $N$ is the number of triplets, $T_i$ and $T_j$ correspond to the binaries for the triplets,
$a_i$ is the bias weight used to evaluate the quality of the triplets, 
and $b_{ij}$ is the coefficient that quantifies the compatibility of
two triplets (Fig.~\ref{fig:quboTrk}: $b_{ij}=0$ if they share no hit, $=1$ if there is any conflict, and $=-S_{ij}$
if they share two hits). 
As the coefficients $a_i$ and $b_{ij}$ are 
unitless, the objective QUBO Hamiltonian $O(a,b,T)$ is accordingly unitless.
The coefficient $-S_{ij}$ quantifies
the consistency of the two triplet momenta via the 
curvature and the angle of the two triplets~\cite{Bapst:2019llh}. 
The bias weight $a_i$ has a significant impact on the Hamiltonian
energy landscape, track reconstruction purity,  
and computation speed. It is a function of the transverse and longitudinal 
displacements of the triplets from the primary vertex 
(the most significant proton-proton collision point of an event), 
optimized with tunable parameters~\cite{Bapst:2019llh}.
The overall workflow of global track finding is schematically described in 
Fig.~\ref{fig:trk_workflow}. The ``solving'' part could be pursued with any 
other Ising problem solvers, such as VQE, QAOA, or quantum-inspired algorithms. 
This study has also adopted the TrackML Challenge dataset and evaluated up to 
6,600 tracks. D-Wave 2X hardware and D-Wave Neal SA provided comparable performance; 
efficiency remained relatively stable, whereas purity tends to degrade against 
the track multiplicity. 

\begin{figure}[t]
    \centering
      \begin{subfigure}{0.9\linewidth}
            \includegraphics[width=\linewidth]{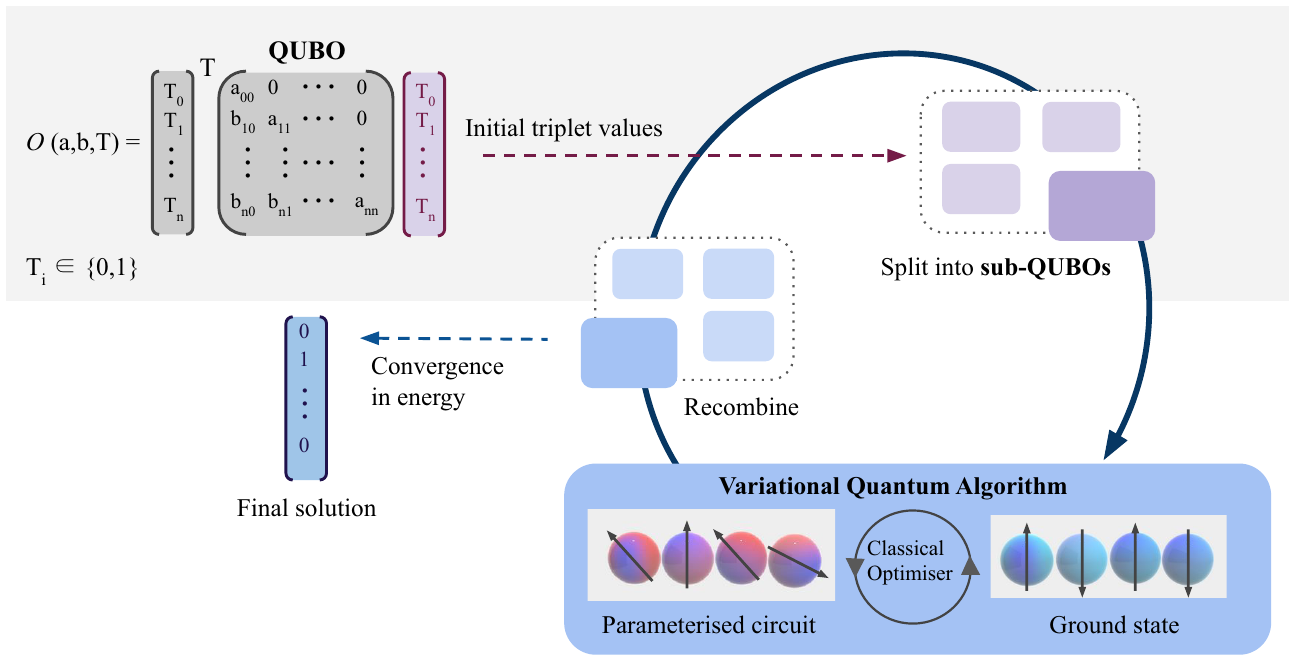}
      \end{subfigure}
      \caption{\label{fig:subQUBO} Schematic diagram illustrating the sub-QUBO procedure. Reproduced from Ref.~\citenum{Crippa:2023ieq}. \href{https://creativecommons.org/licenses/by/4.0/}{CC BY 4.0}.} 
\end{figure}

A notable difference in the approach between Refs.~\refcite{Zlokapa:2019tkn} and \refcite{Bapst:2019llh}
is that the Ising problem was split into sub-matrices and the optimization is iteratively pursued
for the latter, in order to handle the high track multiplicity events.
This approach is called the sub-QUBO method (Fig.~\ref{fig:subQUBO}). 
The track multiplicity is exceedingly high at the HL-LHC, leading to about 0.1M spins/binaries or 
more in the Ising/QUBO formulation. Such large problems 
obviously do not fit in existing quantum circuits and annealing hardware.
D-Wave provides the qbsolv sub-QUBO solver, 
which has been used in Refs.~\refcite{Bapst:2019llh,Funcke:2022dws,Crippa:2023ieq,Schwagerl:2023elf}, 
for example. Refs.~\refcite{Funcke:2022dws,Crippa:2023ieq} performed track reconstruction on 
a simulated dataset from the LUXE experiment~\cite{Abramowicz:2021zja}, where qbsolv was 
used along with VQE. Its track reconstruction performance was comparable to the traditional 
combined Kalman filter. In Ref.~\refcite{Schwagerl:2023elf},
the dependence of the tracking performance on the sub-QUBO matrix sizes was evaluated.
It is worth noting that most sub-QUBO approaches, including qbsolv, are heuristic and lack a
theoretical foundation. Ref.~\refcite{okawa} adopted, for the first time in HEP, 
a theoretically robust sub-QUBO algorithm using multiple solution instances~\cite{subQUBO}. 
The study also introduced QAOA for the first time in track finding~\cite{okawa}, and achieved 
comparable performance to the D-Wave results~\cite{Bapst:2019llh} using 6-qubit Origin 
Quantum Wuyuan hardware.  

\begin{figure}[t]
    \centering
    \begin{subfigure}{0.49\linewidth}
	\includegraphics[width=\linewidth]{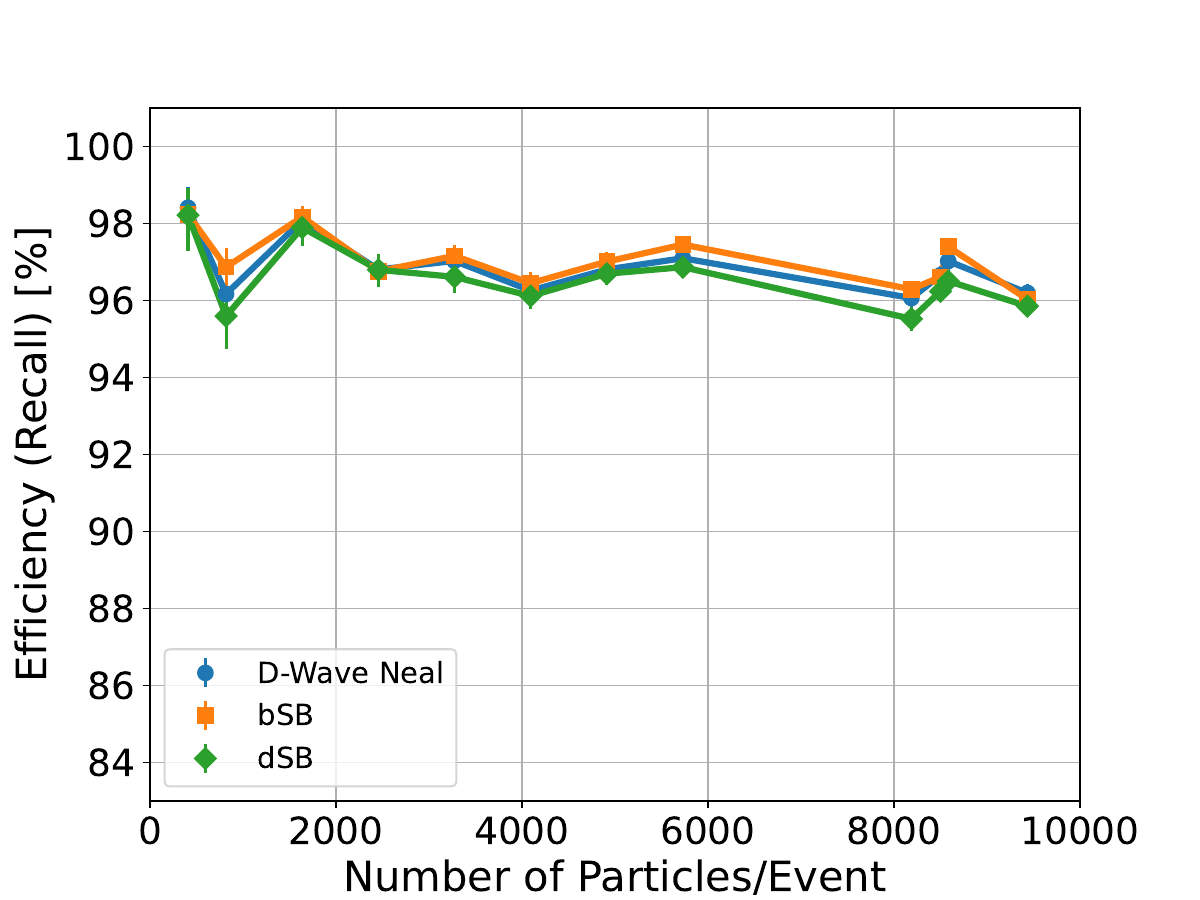}
	\caption{}
    \label{subfig:eff}
    \end{subfigure}
    \begin{subfigure}{0.49\linewidth}
	\includegraphics[width=\linewidth]{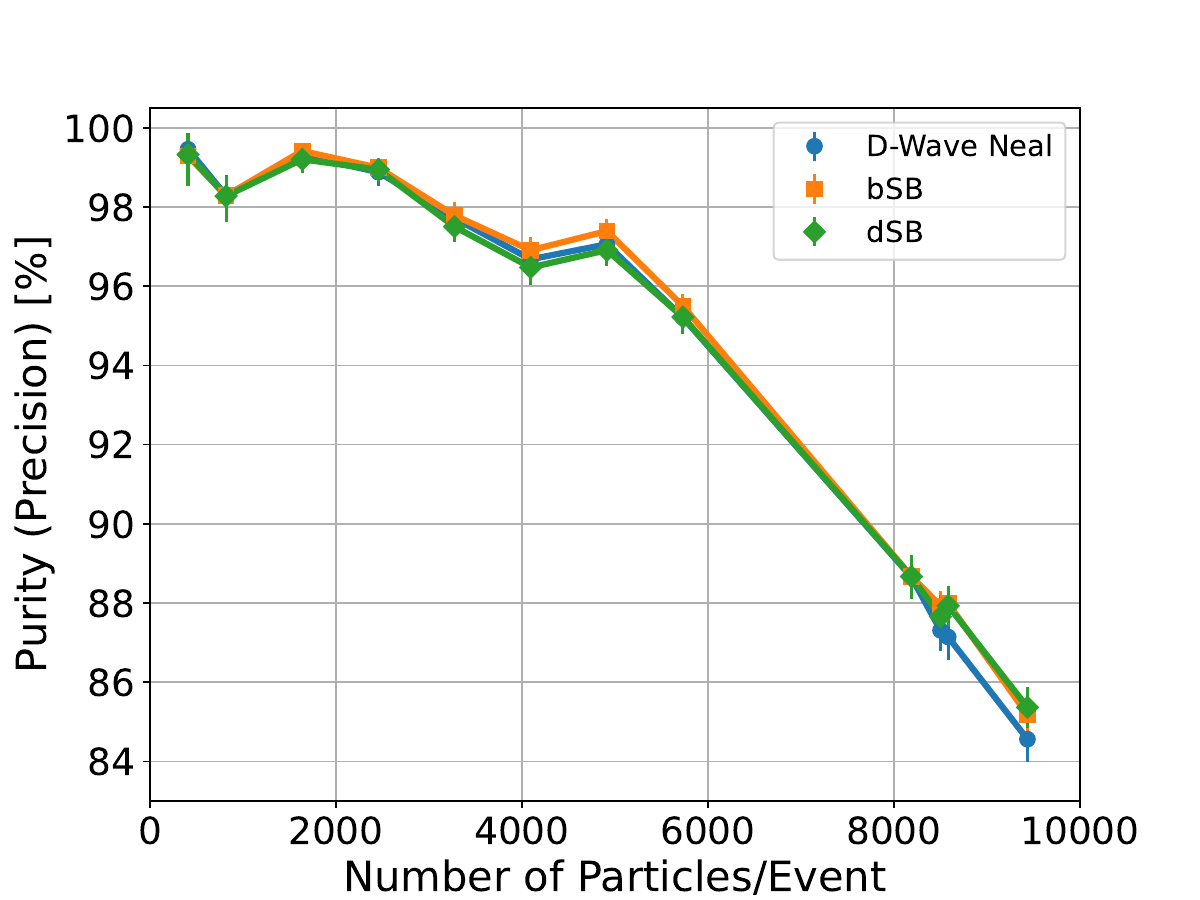}   
    \caption{}
    \label{subfig:purity}
    \end{subfigure}
    \begin{subfigure}{0.49\linewidth}
	\includegraphics[width=\linewidth]{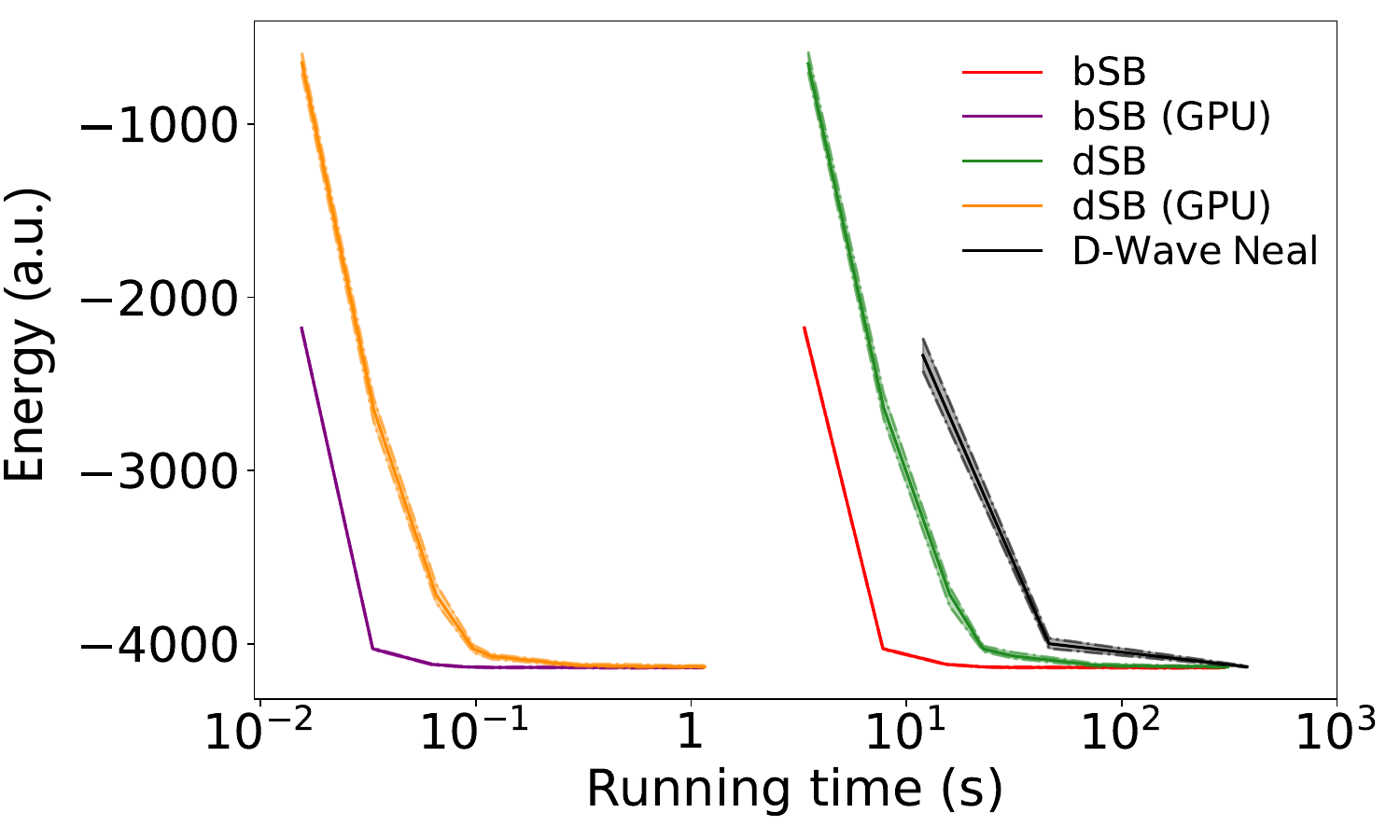}   
    \caption{}
    \label{subfig:time}
    \end{subfigure}
	\caption{Efficiency (recall) (a) and purity (precision) (b) in relation to particle multiplicity, 
    and evolution of Ising energies over time for the event with the highest particle multiplicity 
    (c) evaluated for the three quantum-inspired algorithms. 
    Reproduced from Ref.~\citenum{qaiatrack}. \href{https://creativecommons.org/licenses/by/4.0/}{CC BY 4.0}.}
        \label{fig:track_perf}
  \end{figure}

The tracking studies using quantum annealing or circuit hardware all obtained performance 
comparable to the traditional methods. However, the critical bottleneck 
from the limited hardware resources remains. The sub-QUBO approach 
generally does not lead to degradation in the Ising solving precision itself, but the computational 
speed degrades by two orders of magnitude~\cite{Bapst:2019llh} or more, depending on the 
fineness of the sub-QUBO splitting. In order to overcome this challenge, 
Ref.~\refcite{qaiatrack} investigated quantum-inspired algorithms, bSB, dSB, and D-Wave Neal 
SA in particular, 
as they can directly handle the HL-LHC dataset. Figs.~\ref{subfig:eff} and \ref{subfig:purity}
show the track reconstruction performance in terms of efficiency and purity. 
The SB algorithms provide comparable or slightly better efficiency and purity than D-Wave Neal SA.
Using a single GPU, bSB achieves four orders of magnitude speed-up from D-Wave Neal SA (Fig.~\ref{subfig:time}): 
the reduction from 23 minutes down to 0.14s, for the largest TrackML dataset of 9,435 particle 
multiplicity (Fig.~\ref{fig:trk} shows its event display). 
SB can effectively run multiple processing tasks with cutting-edge computing 
resources such as GPU and FPGA, demonstrating its potential and readiness 
for practical applications. 

A distinctive approach is to regard track reconstruction 
as a classification of signal and background events from the 
detector hits~\cite{Quiroz:2020jmp}, formulated as an optimization 
problem. 
The classification can be pursued with the
quantum associative memory model (QAMM)~\cite{QAMM1,QAMM2} or the quantum content 
addressable memory model (QCAM)~\cite{QCAM} 
on the D-Wave 2000Q quantum annealer. 
The study considered two classification methods: the first
based on the solution state energy and the second based 
on the key bit value for the classification label. 
The optimal choice of the memory and classification method 
depends on the track density. 
The performance is highly sensitive to the pattern density
$\alpha$ and degrades for large densities.

\subsection{Global track finding with neural networks or kernels}
\label{sec:globalTrk_GNN}

As mentioned above, classical GNNs have been actively investigated at various 
collider experiments. 
In the GNN approach, detector hits are regarded as nodes and 
connected track segments as edges. The empirical evidence shows that the 
computing time scales linearly with the number of space points~\cite{ExaTrkX:2021abe}. 
This feature is in clear contrast to the exponential increase observed in 
the combined Kalman filter.  

\begin{figure}[t]
    \centering
    \includegraphics[width=0.9\linewidth]{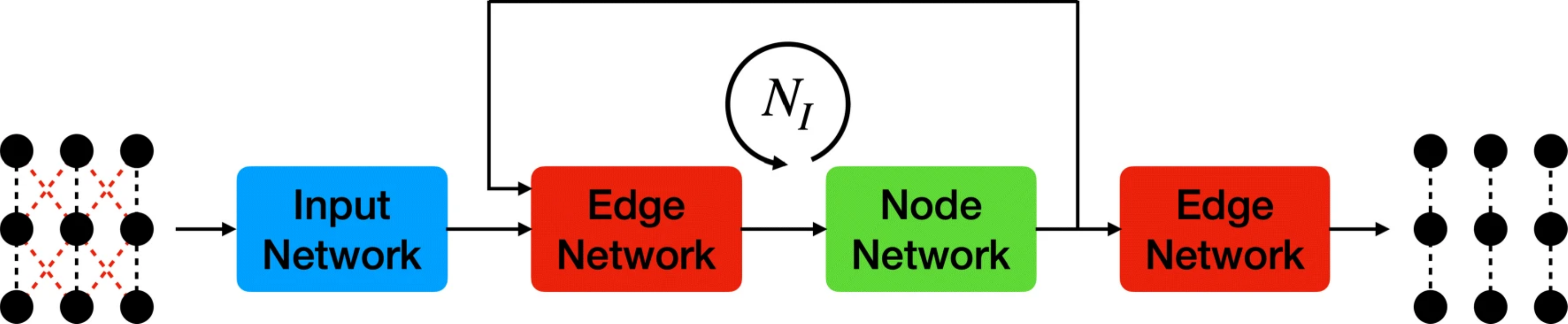}
    \caption{Schematic presentation of the QGNN architecture. 
    Reproduced from Ref.~\citenum{gnn-gate}. \href{https://creativecommons.org/licenses/by/4.0/}{CC BY 4.0}.\label{fig:qgnn}}    
\end{figure}

\begin{figure}[t]
    \centering
    \includegraphics[width=0.9\linewidth]{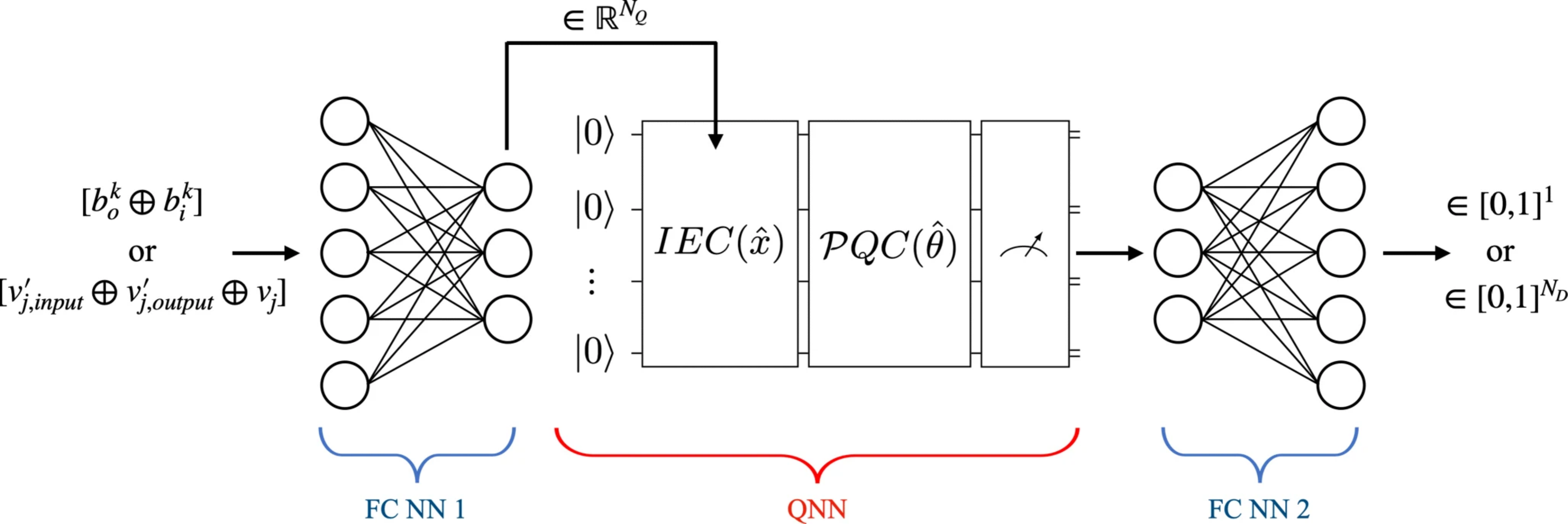}
    \caption{The Hybrid Neural Network architecture consists of classical and quantum layers. 
    Reproduced from Ref.~\citenum{gnn-gate}. \href{https://creativecommons.org/licenses/by/4.0/}{CC BY 4.0}.\label{fig:hnn}}    
\end{figure}

To explore potential benefits from variational quantum layers, 
a hybrid quantum-classical GNN (QGNN) was developed for track reconstruction~\cite{gnn-gate}. 
The QGNN consists of three networks as schematically 
presented in Fig.~\ref{fig:qgnn}.
The Input Network 
accepts the spatial coordinates of the detector hits to a fully connected
neural network layer to form the initial node feature vector. Then the 
node feature vector is passed to Edge and Node Networks that iteratively 
process the graph to determine the final edge probabilities. 
In those networks, both classical and quantum layers are combined to form 
Hybrid Neural Networks (HNNs) (Fig.~\ref{fig:hnn}). The information 
encoding circuit (IEC) maps classical data to qubits. The parameterized 
quantum circuit (PQC) transforms those states to the Hilbert space. 
Finally, measurements are passed back to the classical layers.
Four different configurations of PQCs are investigated in the study, 
for two of which originate from tensor networks: matrix product state (MPS)
and Tree Tensor Network (TTN).

The study showed that the expressibility and entanglement capacities 
of the quantum circuits affect the training performance. Because of 
quantum hardware limitations, the study was conducted with a quantum 
circuit simulator handling up to 16 qubits~\cite{Rieger:2021poc,gnn-gate}. The QGNN model 
achieves comparable performance to the classical counterpart.
It adopted angle encoding, but more sophisticated 
encoding has the potential to improve the performance and is left for 
future studies. The algorithm was also adopted for a LUXE 
simulation dataset~\cite{Crippa:2023ieq} via a quantum circuit simulator, 
achieving a comparable efficiency to the traditional combined Kalman 
filter and the VQE algorithm, but with a visibly higher fake rate. 
Evaluating the 
impact of hardware noise and performing the algorithm on quantum 
hardware is left for future studies. A QUBO version 
of GNN tracking was investigated in Ref.~\refcite{gnn-aneal}.

An alternative approach was proposed to classify triplets as being 
associated with the same track or not, based on 
a support vector machine (SVM) with a quantum-estimated kernel~\cite{PhysRevD.109.052002}. 
The quantum-kernel approach outperforms its classical counterpart for 
datasets with many features. However, specifically for triplet-based 
cases, the quantum-enhanced method did not bring in any benefit. 
 
\subsection{Iterative quantum tracking}

Iterative track reconstruction requires Associative Memory to store 
a database that encompasses hit patterns in the detector. 
The software prototype of Quantum Associative Memory (QuAM)~\cite{Ventura:1998nr,EZHOV2000271,Trugenberger:2001ywy}
using a quantum circuit was discussed for 
data triggering at the LHC~\cite{Shapoval:2019txi}.  
An exponential capacity increase with QuAM was conceptually 
demonstrated, but its practical implementations in IBM 5Q and 14Q 
quantum hardware faced some topological bottlenecks because of the 
limited number of qubits and their connectivity.  

The whole reconstruction chain on a realistic dataset 
is presently unfeasible to pursue 
on quantum hardware; however, its computational complexity was 
evaluated for each step~\cite{Magano:2021jzd}: seeding, track 
building, 
cleaning, and final selection. In contrast to 
the majority of the global reconstruction studies that concentrate 
on the track building
step, this study outlines the complexity of algorithms at each tracking stage,
as summarized in Table~\ref{tab:track_summary}. The algorithm used
at the CMS experiment~\cite{CMS:2014pgm} was adopted as the classical 
benchmark. 
At the seeding stage, we form initial track segments with 
just a few detector hits. Seeding uses the Grover
search algorithm~\cite{PhysRevLett.79.325,Grover2} for the quantum 
approach, leading to the quantum complexity of 
$\tilde{O}\left(\sqrt{k_{\text{seed}} \cdot n^c }\right)$ instead of 
$O\left(n^c\right)$ from the classical algorithm, where 
$k_{\textrm{seed}}$ is the number of seeds, $n$ is the number of charged particles in the events, 
and $c$ is the number of detector hits used for the seeds. 
Then, for track building, those seed trajectories 
are extrapolated along the expected path, adding compatible detector 
hits with the lowest $\chi^2$ from the consecutive detector layers to form 
track candidates. 
The quantum minimum finding algorithm by D{\"u}rr and H{\o}yer~\cite{durr1999}
allows us to find the best track candidates with $\tilde{O}(k_{\text{seed}} \cdot \sqrt{n})$
instead of the classical complexity of $\tilde{O}(k_{\text{seed}} \cdot n)$.
As multiple track candidates could describe the same particle, a cleaning 
process removes track candidates that share the majority of the detector hits.
Finally, only the track candidates satisfying some criteria based on
the trajectory fit quality are kept as the final tracks.   
The study demonstrated that quantum algorithms can reconstruct the 
same tracks as the classical algorithm with lower quantum 
complexity, in particular for seeding and track building. However, 
the expected quantum advantage with this approach is rather mild, 
and the authors speculate on a potentially higher success in designing 
a completely new algorithm, which aligns with the directions 
already mentioned in Sections~\ref{sec:globalTrk_Ising} and \ref{sec:globalTrk_GNN}. 

\begin{table}[t]
\caption{Computational complexity of track reconstruction at each stage with classical and quantum algorithms~\cite{Magano:2021jzd}. 
$k_{\textrm{cand}}$ is the number of track candidates. The original version of the cleaning stage refers to the 
method used by the CMS experiment~\cite{CMS:2014pgm}, whereas the authors of this study found an algorithm with lower 
complexity~\cite{Magano:2021jzd}.   
\label{tab:track_summary}}
\resizebox{\textwidth}{!}{%
\begin{tabular}{ccccc}
\hline\hline
\textbf{Tracking stages}      & \textbf{Input size}    & \textbf{Output size}& \textbf{Classical complexity}    & \textbf{Quantum complexity}  \\ \hline\hline
\textbf{Seeding}              & $O(n)$  & $k_{\text{seed}}$& $O\left(n^c\right)$ & $\tilde{O}\left(\sqrt{k_{\text{seed}} \cdot n^c }\right)$   \\ \hline
\textbf{Track Building}       & $k_{\text{seed}}+ O(n)$ & $k_{\text{cand}}$& $O(k_{\text{seed}} \cdot n)$  & $\tilde{O}\left(k_{\text{seed}} \cdot \sqrt{n}\right)$  \\ \hline
\textbf{Cleaning (original)}  &  $k_{\text{cand}}$ & $O(k_{\text{cand}})$& $O(k_{\text{cand}}^2)$   & --     \\ \hline
\textbf{Cleaning (improved)}  &  $k_{\text{cand}}$ & $O(k_{\text{cand}})$& $\tilde{O}(k_{\text{cand}})$  & --  \\ \hline
\textbf{Selection}            &  $O(k_{\text{cand}})$ & $O(k_{\text{cand}})$& $O(k_{\text{cand}})$  & -- \\ \hline
\textbf{Full Reconstruction}  & $n$  & $O(n^c)$& $O\left(n^{c + 1}\right)$  & $\tilde{O}\left(n^{c + 0.5}\right)$   \\ \hline
\textbf{\begin{tabular}[c]{@{}c@{}}Full Reconstruction with\\  
    $O(n)$ reconstructed tracks\end{tabular}} & $n$   
    & $O(n)$& $O\left(n^{c + 1}\right)$
    & $\tilde{O}\left(n^{(c + 3)/2}\right)$  \\ \hline\hline
\end{tabular}%
}
\end{table}

\begin{figure}[t]
    \centering
    \includegraphics[width=0.8\linewidth]{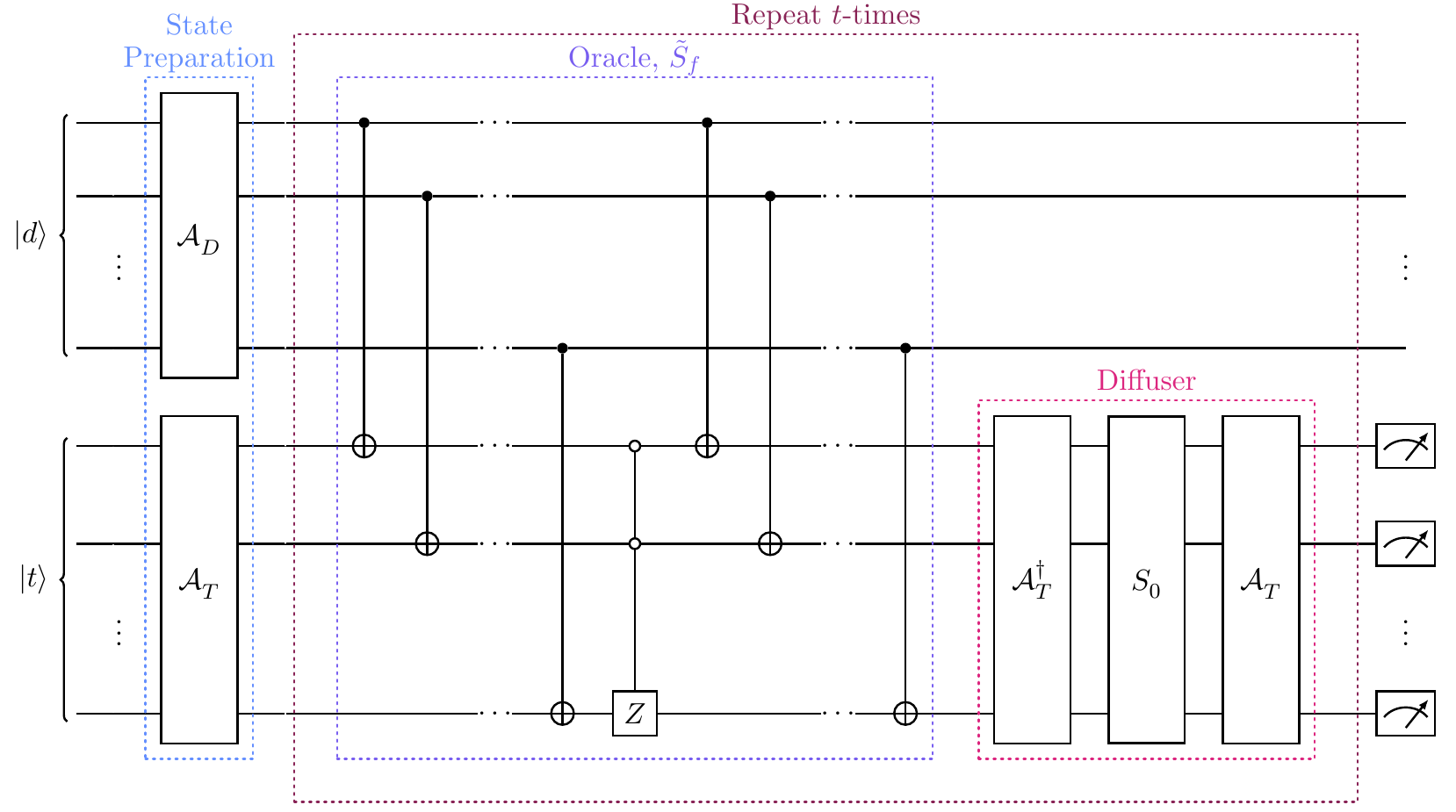}
    \caption{Schematic circuit diagram of Quantum Amplitude Amplification (QAA). 
    Reproduced from Ref.~\citenum{Brown:2023llg}. \href{https://creativecommons.org/licenses/by/4.0/}{CC BY 4.0}.\label{fig:qaa}}    
\end{figure}

An alternative iterative approach using a novel oracle design 
was proposed in Ref.~\refcite{Brown:2023llg}, 
which extends the idea of using QuAM~\cite{Shapoval:2019txi}.
The study considered a simplified detector model with four layers of 
three tracking modules and used a quantum circuit simulator to 
demonstrate the performance. Track reconstruction with a quantum version of 
the template matching algorithm~\cite{Bardi:1997hm,Nicolaidou:2010zz,Bunkowski:2019gac,ATLAS:2021tfo} 
was considered. The study adopted a generalization of the Grover 
search algorithm: Quantum Amplitude Amplification (QAA)~\cite{Brassard:2000xvp},
which improves the efficiency of database encoding. QAA 
consists of two components: the oracle, which labels the states of interest by 
inverting their phase, and the diffuser, which enlarges the amplitudes of the 
marked states (Fig.~\ref{fig:qaa}). To match the 12-module simplified detector, the study used 
12 qubits to perform QAA.  
As the Quantum Random-Access Memory (QRAM)~\cite{Giovannetti:2007ope}
is not available, the database is prepared via the state preparation routine in IBM Qiskit. It uses 
a recursive initialization algorithm with optimization~\cite{Shende:2006onn}. 
The quantum circuit consists of the data and template registers, 
following the oracle construction from Ref.~\refcite{Gao:2021rxg}.
The detector hits are encoded onto the data register, and the template  
register encodes the QAA routine with the unitary 
operations $\mathcal{A}_D$ and $\mathcal{A}_T$, respectively. For the simplified 
dataset under consideration, 15 possible patterns exist and are one-hot encoded into bit 
strings of 12 qubits, with each bit corresponding to the detector module. 
When a module detects a hit, its bit is flipped to ``1''; otherwise, it remains as
``0''. 
Then the general oracle marks the state in the template database that corresponds to the 
detector hit pattern encoded on the data register. The diffuser operators then amplify
the marked amplitudes. With three iterations of the QAA routines and $10^4$ shots on the 
qasm\_simulator without a noise model, this method achieved a track finding efficiency larger 
than 90\% for events without any detector failure. The study also demonstrated that 
the oracle can be modified to account for missing hits from detector failures. 
Notably, this modification does not lead to additional computational complexity,
which is in clear contrast to classical techniques.   

\subsection{Vertex reconstruction}

Primary vertices are positions of beam-particle interactions at high-energy 
colliders and are crucial components to fully reconstruct collision events 
based on the detector information. Primary vertices are reconstructed by 
clustering charged-particle tracks along the beam axis. 
Ref.~\refcite{Das:2019hrw} developed a quantum-annealing approach to 
reconstruct primary vertices from simplified artificial datasets, which 
resemble conditions in the Compact Muon Solenoid (CMS) experiment
at the LHC. 

To cluster the tracks, the QUBO Hamiltonian $Q_{\mathrm{vtx}}$ was defined as: 
\begin{equation}
Q_{\mathrm{vtx}} = \sum_{k}^{n_V} \sum_i^{n_T} \sum_{j>i}^{n_T}
p_{ik} p_{jk} g(D(i,j);m) + \lambda \sum_i^{n_V} \left( 1 - \sum_k^{n_V} p_{ik} \right),
\end{equation}
where $i$ and $j$ are indices for tracks, $k$ is the index for vertices, 
$n_V$ is the number of vertices, $n_T$ is the number of tracks, and $p_{ik}$
is the probability of the $i$-th track to be associated with the $k$-th vertex.
$g$ is a distortion function that measures the deformation between a source and its 
measurement, $m$ is a distortion parameter, 
and $D(i,j)$ is a distance metric between the $i$-th and $j$-th tracks.
$g$ and $D(i,j)$ are defined as: 
\begin{eqnarray}
g(x,m) &=& 1 - e^{-mx}, \\
D(i,j) &=& \frac{|z_i - z_j|}{\sqrt{\delta z_i^2 + \delta z_j^2}},
\end{eqnarray}
where $z_i$ is the longitudinal displacement parameter $z_0$ of the $i$-th track, 
$\delta z_i$ is its measurement uncertainty, and $\lambda$ is the parameter 
for the penalty term that ensures the probability to add up to one. 

The QUBO Hamiltonian was solved with a D-Wave 2000Q quantum annealer for 
various simplified event topologies ranging from two primary vertices and 10 
tracks to five primary vertices and 15 tracks. The performance was comparable 
with simulated annealing for small numbers of vertices. Extending this method 
to larger problem sizes compatible with the LHC and HL-LHC would require an increased
number of qubits and connectivity in quantum annealing hardware. 

\section{Jet reconstruction}
\label{sec:jetreco}

\begin{figure}[t]
    \centering
    \includegraphics[width=1.0\linewidth]{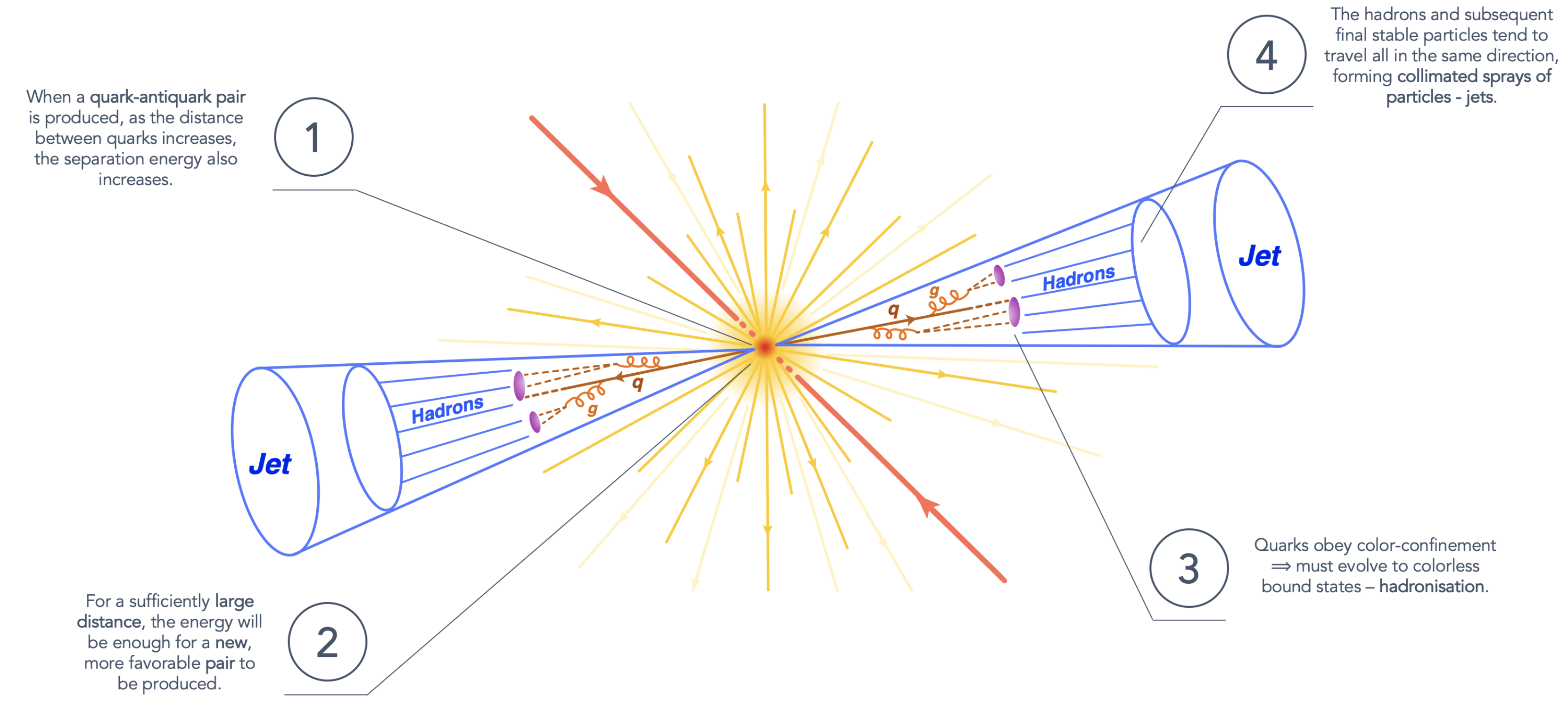}
    \caption{A dijet event from an $e^+e^-$ collision, where a quark-antiquark pair is produced. 
    Reproduced from Ref.~\citenum{digitalqjet}. \href{https://creativecommons.org/publicdomain/zero/1.0/}{CC0 1.0}.\label{fig:jets}}    
\end{figure}

Jet reconstruction is a clustering problem designed to provide reliable 
proxies to determine 
the original quark and gluon kinematics (Fig.~\ref{fig:jets}). 
Due to color confinement, quarks and gluons produced at the high-energy 
colliders or from the decays of heavy particles initiate 
cascades of collimated particles originating from their fragmentation 
and hadronization. 
Jet reconstruction is a sophisticated computational procedure to retrieve
such information. 

The idea of jet clustering dates back to the proposal by Sternman and 
Weinberg~\cite{weinberg}. Since then, various algorithms have been developed 
over decades, as reviewed in Refs.~\refcite{jetrev1,jetrev2,jetrev3,jetrev4,jetrev5}.
Traditional jet clustering algorithms are dominantly based on iterative approaches,
of which the most standard class is the sequential recombination algorithms.
The closest pair of jet inputs, such as tracks, calorimeter clusters, or particle 
flow objects, are repeatedly recombined until the algorithm terminates by 
all inputs being clustered within a user-defined distance $R$ or 
a user-defined jet multiplicity. The former is called inclusive clustering and 
is used in hadron collider experiments, whereas the latter is called 
exclusive clustering and is adopted at electron-positron colliders. 
Just as track reconstruction, quantum algorithms for jet clustering have 
been investigated with two approaches: global or iterative.  

\subsection{Iterative jet reconstruction}
\label{sec:itrjet}

The first pioneering study evaluated the computational complexity of
finding the thrust axis to separate each event into two hemispheres 
for dijet events~\cite{thaler1}. The authors demonstrated how to formulate
thrust reconstruction as a partitioning problem for quantum annealing or an
axis-finding problem to be pursued by Grover search.  
The former is a global reconstruction approach, which is described 
in Section~\ref{sec:Gjet}. As the axis-finding problem, 
thrust $T$ is defined as 
finding the optimal unit vector that maximizes $T(\hat{n})$: 
\begin{eqnarray}
\label{eq:thrust_axis}
T(\hat{n})&=&\frac{\sum_{i=1}^N|\hat{n}\cdot\vec{p}_i|}{\sum_{i=1}^N|\vec{p}_i|}, \\
T &=& \max_{|\hat{n}|=1} T(\hat{n}), 
\end{eqnarray}
where $\hat{n}$ is a unit norm vector to scan and $\vec{p}_i$ is the three 
momentum of the $i$-th particle. 
Table~\ref{tab:thaler1} summarizes the computational scaling for $N$ particles
in the final state.  The most efficient classical algorithm previously known 
as of Ref.~\refcite{thaler1} scales as $O(N^3)$~\cite{Yamamoto:1983bfq}. 
The authors first of all proposed that the classical algorithm can be 
improved to $O(N^2 \log N)$ with a sorting technique inspired by 
the \textsc{SISCone} jet algorithm~\cite{Salam:2007xv}. 
By introducing 
parallel computing models in which $N$ parallel processors accompany 
$N$ words of memory, the runtime can be further reduced to  
$O(N \log N)$. For quantum approaches, the authors adopted the quantum 
maximum finding algorithm of D\"urr and H\o{}yer~\cite{Durr:1996nx}, 
which is a generalization of Grover search. For loading classical datasets 
into a quantum memory, they studied two computing scenarios: the sequential 
model and the parallel model.  In the sequential model, one classical or 
quantum gate is executed per time step, whereas in the parallel model, 
all $N$ database items are preloaded to $O(N)$ qubits at the same time. 
Quantum thrust algorithms generally have two abstract operations: 
LOOKUP and SUM~\cite{thaler1}. The LOOKUP operation extracts the momentum 
of a particle corresponding to one index, whereas the SUM operation adds up
all momenta. In the sequential (parallel) model, both the LOOKUP and SUM 
operations take $O(N)$ ($O(\log N)$) time. The parallel approaches with 
classical and quantum thrust algorithms coincidentally end up with the 
same $O(N\log N)$ scaling, but due to fundamentally different reasons. 

\begin{table}[t]
    \caption{Computational complexity of classical and quantum thrust algorithms 
for a single collision event with $N$ particles~\cite{thaler1}. }
\centering
\begin{tabular}{l @{$\quad$} l @{$\quad$} l }
\hline
\hline
\textbf{Implementation} & \textbf{Time Usage} & \textbf{Qubit Usage} \\
\hline
\hline
Classical~\cite{Yamamoto:1983bfq} & $O(N^3)$ & --- \\
Classical with sort inspired by \textsc{SISCone}~\cite{Salam:2007xv} & $O(N^2 \log N)$ & --- \\
Classical with parallel sort & $O(N \log N)$ & --- \\
\hline
Quantum annealing & Gap dependent & $O(N)$ \\
Quantum search: sequential model & $O(N^2)$ & $O(\log N)$  \\
Quantum search: parallel model & $O(N \log N)$ & $O(N\log N)$ \\
\hline
\hline
\end{tabular}
\label{tab:thaler1}
\end{table}

Subsequent studies used the IBM Qiskit circuit simulator to actually 
pursue iterative quantum jet clustering~\cite{digitalqjet,quantumjet}.
Ref.~\refcite{digitalqjet} adopted the \texttt{K-means} algorithm~\cite{kmeans1,kmeans2} 
to inclusively reconstruct jets, iterating over $N$ jet constituents with 
$D$-dimensional data points and outputting $K$ jets where $K$ is defined 
by the user. As the jet multiplicity is unknown ahead of time in inclusive 
jet reconstruction, the algorithm iterated over a range of $K$ and 
chose the jet multiplicity with the highest Silhouette Index~\cite{ROUSSEEUW198753}.
The computing complexity scales as $O(KND)$ for the classical \texttt{K-means}, 
whereas it is reduced to $O(KN\log D)$ in the quantum counterpart due to the 
data encoding of particle momenta and jet centroids (Table~\ref{tab:deLejarza}). 
Jet reconstruction with quantum \texttt{K-means} resembled the $k_t$ jet 
clustering~\cite{Catani:1993hr} in terms of jet constituent assignment and jet multiplicity. 

\begin{table}[t]
    \caption{Computational complexity of classical and quantum algorithms using 
     \texttt{K-means}, affinity propagation (AP), or sequential 
    recombination algorithms for a single 
     collision event with $N$ particles, $K$ centroids (\texttt{K-means}), $T$ iterations (AP), and 
     dimensions of data points $D$ (\texttt{K-means}, AP)~\cite{digitalqjet,quantumjet}. }
\centering
\resizebox{\textwidth}{!}{%
\begin{tabular}{l @{$\quad$} l @{$\quad$} l }
\hline
\hline
\textbf{Jet algorithms} & \textbf{Classical} & \textbf{Quantum} \\
\hline
\hline
\texttt{K-means} & $O(NKD)$ & $O(NK\log D)$~\cite{digitalqjet}, $O(N\log K\log (D-1))$~\cite{quantumjet}  \\
AP & $O(N^2 TD)$ & $O(N^2 T \log(D-1))$~\cite{quantumjet} \\
$k_t$, C/A, anti-$k_t$ & $O(N^3)$ [suboptimal] & $O(N^2 \log N)$~\cite{quantumjet} \\
                 & $O(N \log N)$ [\texttt{FastJet}]~\cite{Cacciari:2005hq,fastjet} & $O(N \log N)$~\cite{quantumjet} \\
\hline
\hline
\end{tabular}}
\label{tab:deLejarza}
\end{table}

In Ref.~\refcite{quantumjet}, the authors have introduced two novel approaches 
to enhance computational performance: a quantum subroutine to calculate a 
Minkowski-based distance between two data points and another subroutine to 
find the maximum from an unsorted data list. The performance of the quantum 
circuit was evaluated against three sets of classical benchmarks, respectively, 
consisting of \texttt{K-means}, affinity propagation (AP), and sequential 
recombination algorithms. The sequential recombination algorithms generally 
cluster jet constituents based on a distance measure. 
The Minkowski-type quantum distance was 
considered for the first time with the \texttt{K-means} and AP algorithms. 
Computing the distance would only require $O(\log(D-1))$ qubits for both 
the \texttt{K-means} and AP algorithms, and searching for the minimum 
distance with respect to the centroids would yield to $O(\log K)$ for 
\texttt{K-means}. Overall, the quantum \texttt{K-means} and AP algorithms 
achieve reduced computational complexity from the classical counterparts 
(Table~\ref{tab:deLejarza}).
For the sequential recombination clustering, 
the $k_t$~\cite{Catani:1993hr,Ellis:1993tq,Dasgupta:2002bw}, 
Cambridge/Aachen~\cite{Dokshitzer:1997in,Wobisch:1998wt}, 
and anti-$k_t$ algorithms~\cite{Cacciari:2008gp} are evaluated, for which the 
distance measure $d_{ij}$ is defined as: 
\begin{eqnarray}
d_{ij} &=& \min(p_{T,i}^{2p},p_{T,j}^{2p}) \Delta R_{ij}^2 / R^2, \\
\Delta R_{ij} &=& (y_i - y_j)^2 + (\phi_i - \phi_j)^2,
\end{eqnarray}
where $i,j$ are the jet constituent labels; $p_{T,i}$, $y_i$, $\phi_i$ 
are the transverse momentum, rapidity, and azimuthal angle of the 
$i$-th jet constituent; and $R$ is the jet-radius parameter usually taken  
to be between 0.4 and 1. The index parameters $p$=1 ($k_t$), 0 (Cambridge/Aachen), 
and $-1$ (anti-$k_t$) correspond to the three algorithms, respectively. 
The classical sequential recombination 
originally scaled as $O(N^3)$, but later improved to $O(N^2)$ by 
identifying particles' geometrical nearest neighbors, and even down 
to $O(N\log N)$ by further introducing the Voronoi 
diagrams~\cite{Cacciari:2005hq,fastjet}. The quantum circuit
brings in speed-up for the minimum search, scaling the computing 
complexity from $O(N)$ of the classical counterpart down 
to $O(N\log N)$. In summary, the most efficient classical and quantum 
algorithms for the sequential recombination lead to the same order. 
Pursuing the computation on actual quantum hardware was not feasible 
because of the hardware noise and the lack of QuAM. Noise-free 
quantum circuit simulator from IBM demonstrated that the classical 
and quantum algorithms obtain consistent particle assignment to jets.

\subsection{Global jet reconstruction}
\label{sec:Gjet}

Jet reconstruction can also be formulated as a QUBO problem. The first study
on global jet clustering using quantum annealing considered a
thrust-based QUBO Hamiltonian:
\begin{equation}
	O_{\textrm{QUBO}}({s_i}) = \left( \sum_{i=1}^{N} |\vec{p}_i|\right)^2 T({s_i})^2,
	\label{eq:qubo_thrust}
\end{equation}
where $s_i$ is the binary \{0,1\} that indicates which jet the $i$-th constituent is 
assigned to, $N$ is the number of jet constituents,
$\vec{p}_i$ is the three momentum of the $i$-th constituent, and $T({s_i})$ is thrust defined as: 
\begin{equation}
	T({s_i}) = 2 \frac{\left|\sum_{i=1}^{N} s_i \vec{p}_i \right|}{\sum_{i=1}^{N} \left|\vec{p}_i \right|}.
	\label{eq:thrust}
\end{equation}
As described in Section~\ref{sec:itrjet}, global reconstruction of thrust is 
defined as a partitioning problem, and thrust is defined in a slightly different 
formulation as the axis finding problem for the iterative reconstruction 
(Eq.~\ref{eq:thrust_axis}). However, they are actually equivalent quantities.  
The performance was evaluated on the $e^+e^- \rightarrow \gamma/Z^0 \rightarrow q\bar{q}$ process 
with a 5000+-qubit quantum annealer D-Wave Advantage 4.1. After tuning the annealing parameters, namely, 
the relative chain strength, annealing time, and the number of runs per event, they observed 
deviations in the range of 1 to 3 \% from the target value, whereas
the results degraded to the deviations of 50\% without the tuning. 
In summary, quantum annealing with tuned annealing parameters and 
a hybrid quantum/classical approach using the untuned annealing results 
as a seed for classical iterative improvement,
both exhibit similar performance to exact classical approaches and heuristics. 

\begin{figure}[t]
    \centering
    \begin{subfigure}{0.49\linewidth}
    	\includegraphics[width=\linewidth]{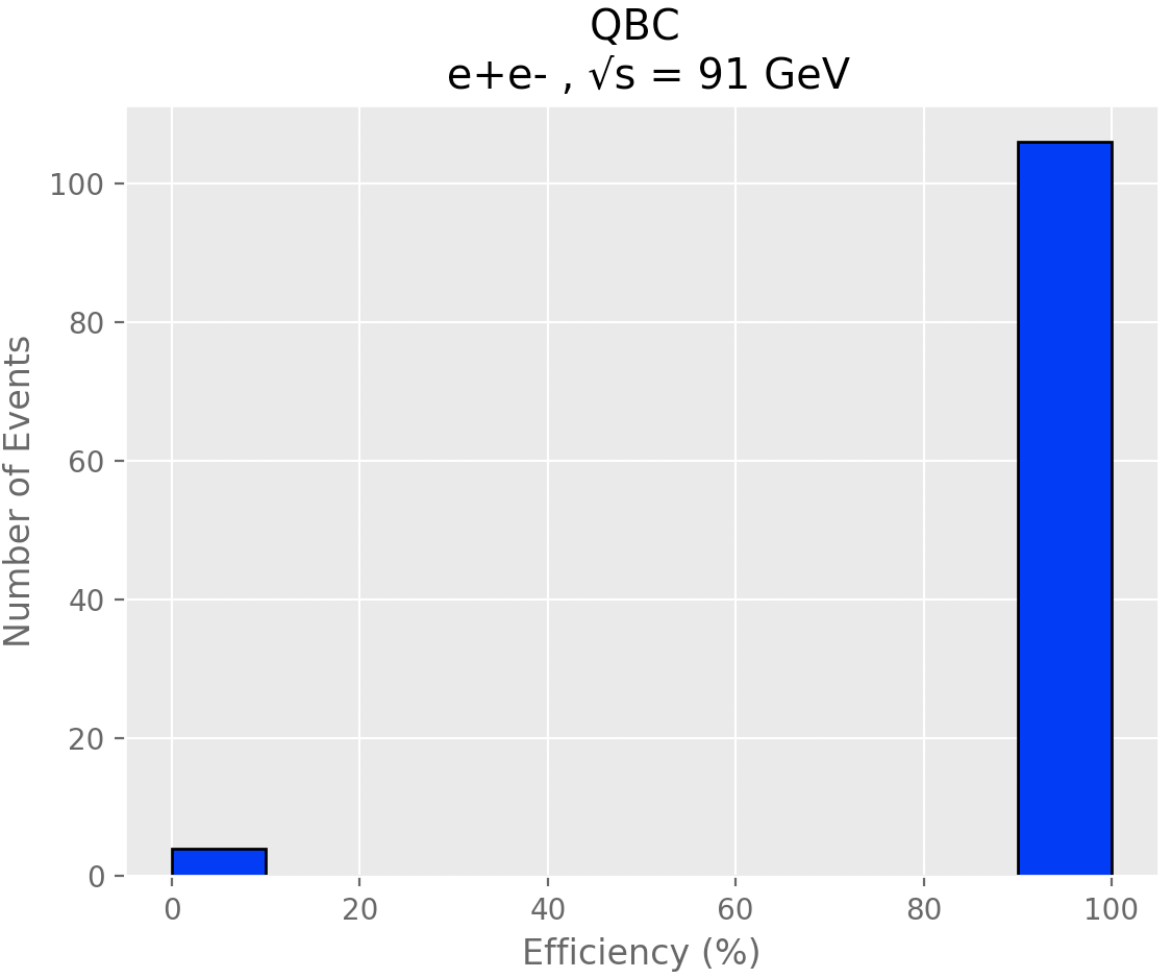}
        \caption{}
        \label{subfig:display_bSB}
    \end{subfigure}
    \begin{subfigure}{0.49\linewidth}
	\includegraphics[width=\linewidth]{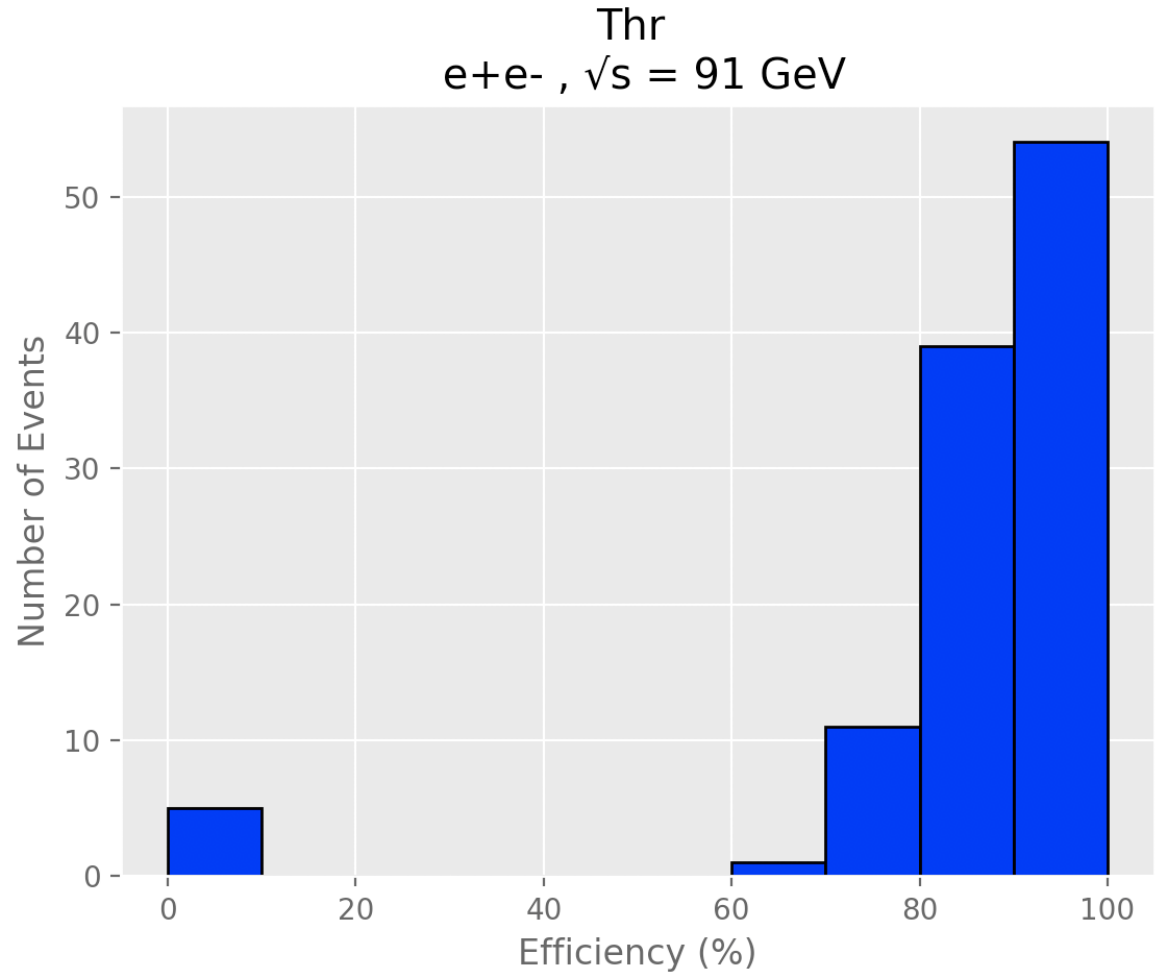}	
        \caption{}
        \label{subfig:mt}
	\end{subfigure}
	\caption{Efficiency of jet clustering with (a) quantum angle-based algorithm and (b) the thrust-based algorithm. 
    Reproduced from Ref.~\citenum{adiabaticjet}. \href{https://creativecommons.org/licenses/by/4.0/}{CC BY 4.0}.
    \label{fig:adiabaticjet} }
\end{figure}

An alternative quantum-annealing approach is to use a more generic formulation of 
the QUBO Hamiltonian~\cite{adiabaticjet} defined as in Eq.~\ref{eq:qubo}. 
In Ref.~\refcite{adiabaticjet},
the angle between the constituents is adopted as the distance: 
\begin{equation}
Q_{ij} = -\frac{\vec{p}_i \cdot \vec{p}_j}{2 (|\vec{p}_i| \cdot |\vec{p}_j|)}.
\label{eq:angle}
\end{equation}
The efficiency defined below was used as the metric to evaluate the jet 
clustering performance:   
\begin{equation}
\textrm{Efficiency} = \frac{\textrm{\# of constituents clustered the same as $ee$-$k_t$}}
            {\textrm{\# of constituents from $ee$-$k_t$}}.
            \label{eq:eff}
\end{equation}
As presented in Fig.~\ref{fig:adiabaticjet}, the angle-based dijet clustering 
is found to cluster jet constituents in a more consistent way than 
the thrust-based approach to the traditional $ee$-$k_t$ algorithm~\cite{eekt}.

A similar study using QAOA as the Ising solver was presented in Ref.~\refcite{zhouchenjet}.
It adopted the sum of the two reconstructed jets as the performance metric and was 
compared to the classical $ee$-$k_t$ and \texttt{K-means} algorithms.
This angle-based approach with QAOA presented comparable performance to the $ee$-$k_t$ algorithm
on simplified 6-particle and 30-particle datasets from the 
$e^+ e^- \rightarrow ZH \rightarrow \nu\bar{\nu} s \bar{s}$ 
events at CEPC. 
Only the former dataset designed for six qubits 
was considered for quantum hardware computation  
because of the noise and limited connectivity of the BAQIS Quafu 
Baihua quantum processor.
The classical \texttt{K-means} algorithm generally presented a
degraded performance compared to the classical $ee$-$k_t$ algorithm and QAOA results. 

As is evident from the binary formulation, all the above formalisms can only handle 
dijet clustering. 
In order to expand the method to multijet problems, 
the QUBO can be generalized to~\cite{thaler1,adiabaticjet,qaiajet}: 
\begin{equation}
	O_{\textrm{QUBO}}^{\textrm{multijet}}({s_i^{(n)}}) = \sum_{n=1}^{n_{\textrm{jet}}}
	\sum_{i,j=1}^{N} Q_{ij} s_i^{(n)} s_j^{(n)}
	 + \lambda \sum_{i=1}^{N} 
	 \left( 1 - \sum_{n=1}^{n_{\textrm{jet}}} s_i^{(n)} \right)^2,
	\label{eq:quboMJ}
\end{equation}
where $n$ considers the jet multiplicity and the binary $s_i^{(n)}$ is 
defined for each jet. The second term serves as a constraint to
ensure that each jet constituent is assigned to a jet only once. 
The coefficient of this penalty $\lambda$ must be sufficiently large: 
$\lambda > N~\textrm{max}_{ij} Q_{ij}$~\cite{thaler1}.
Quantum annealing was explored for multijet 
reconstruction; however, employing multiple qubits for one-hot 
encoding is challenging and susceptible to errors~\cite{adiabaticjet}. Furthermore, 
annealing time tends to have a long duration. It remains to be seen with next-generation 
annealers with the increased qubit connectivity, whether we can address multijet 
reconstruction problems with actual quantum hardware. 

Meanwhile, a quantum-inspired approach using bSB succeeded in globally reconstructing multijet 
events with a QUBO formulation for the first time (Figure~\ref{subfig:display_bSB})~\cite{qaiajet}.  
The study considered $e^+ e^- \rightarrow Z \rightarrow q\bar{q}$, $ZH \rightarrow q\bar{q}b \bar{b}$,
and $t\bar{t} \rightarrow bq\bar{q} \bar{b}\bar{q}q$ events at CEPC, corresponding to 
2, 4, and 6 jets, respectively. 
The distance used in the Durham or \eekt{} algorithm~\cite{eekt} was adopted for the 
QUBO matrix: 
\begin{equation}
	Q_{ij} = 2 \textrm{min}(E_i^2,E_j^2)(1-\cos\theta_{ij}),
\end{equation}
where $E_i$ is the energy of the $i$-th jet constituent, while $\theta_{ij}$
denotes the angle between the $i$-th and $j$-th jet constituents. 
For the all-hadronic $t\bar{t}$ events, bSB was able to find about 10 times lower minimum energy
prediction compared to SA. The study also showed that the angle-based formulation is 
not suitable for high jet multiplicity events, and thus, the QUBO formulation has a 
significant impact on the multijet clustering performance. 
Furthermore, global reconstruction by bSB improved the jet energy resolution 
by 6 (7)\% in the invariant mass of 
Higgs bosons (top quarks) from their hadronic decays.  Ref.~\refcite{qaiajet}
also evaluated the performance of bSB against quantum annealing and QAOA 
for two simplified jet datasets (12 qubits) using simulators~\cite{morrell2024quantumannealing,mindquantumpaper}. 
Even for such small datasets, bSB outperforms the computing speed of D-Wave 2000Q by about two orders of 
magnitude and even more for QAOA, although it should be noted that QAOA should run faster 
on real quantum hardware.

\begin{figure}[t]
    \centering
    \begin{subfigure}{0.49\linewidth}
    	\includegraphics[width=\linewidth]{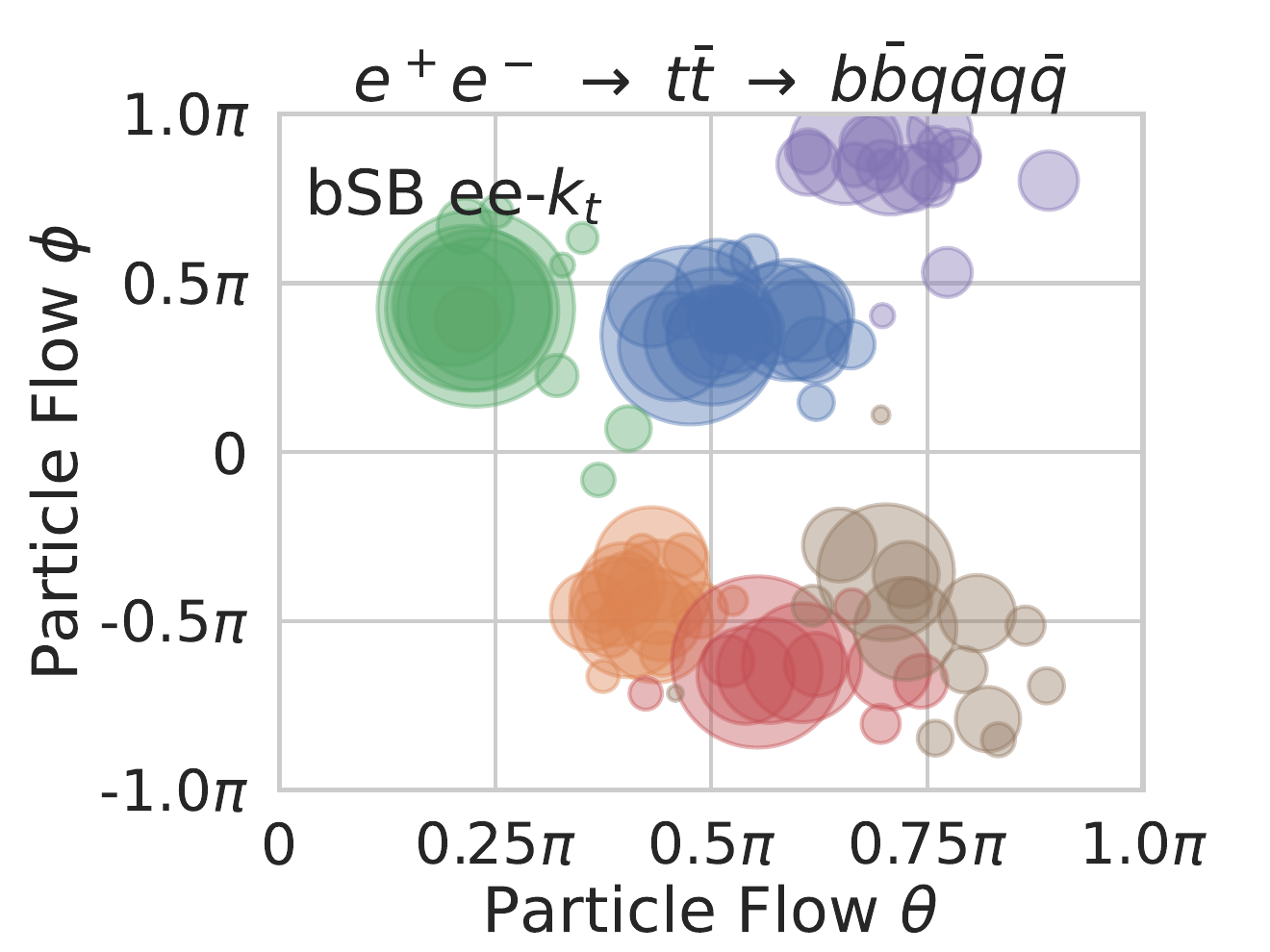}
        \caption{}
        \label{subfig:display_bSB}
    \end{subfigure}
    \begin{subfigure}{0.49\linewidth}
	\includegraphics[width=\linewidth]{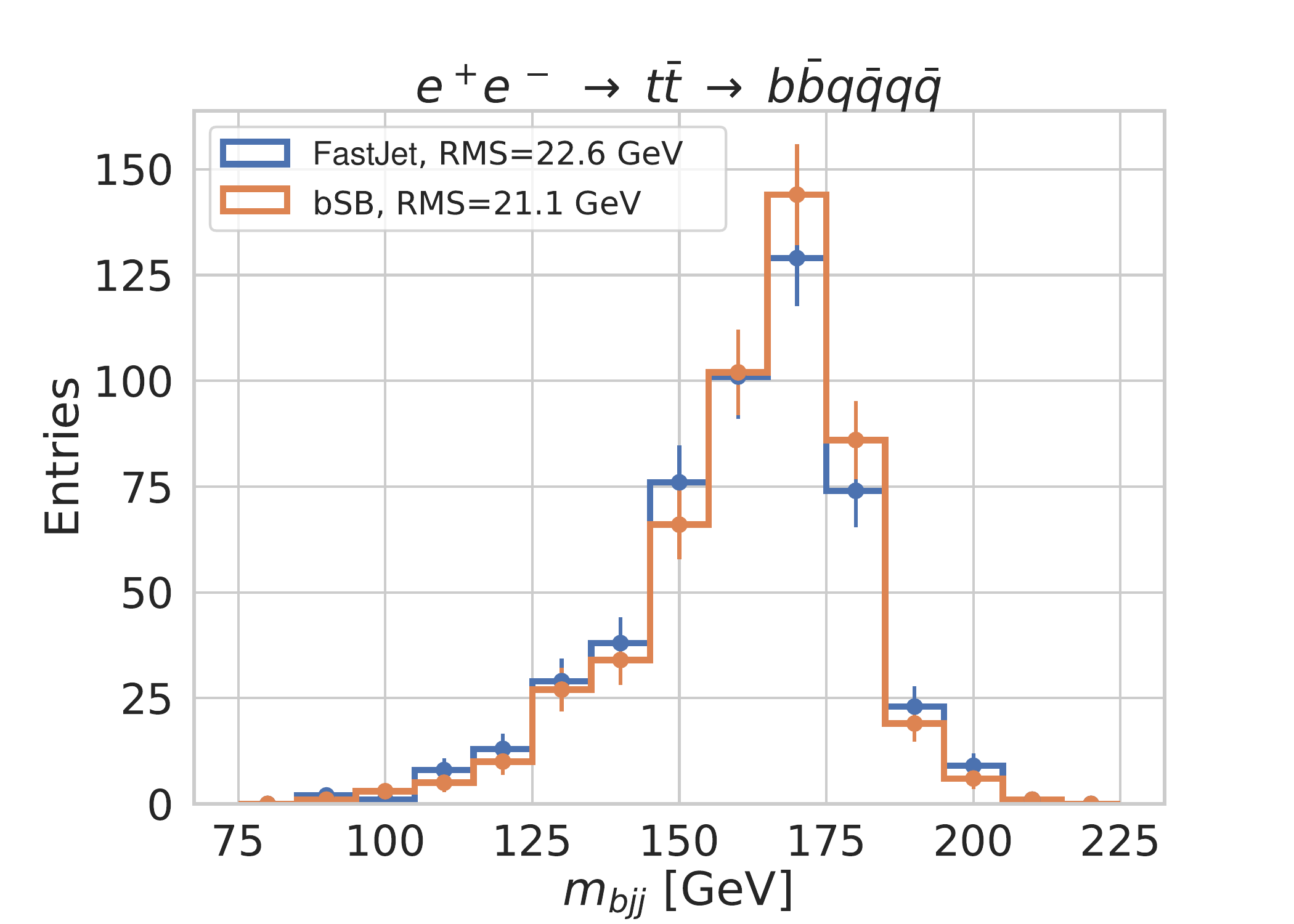}	
        \caption{}
        \label{subfig:mt}
	\end{subfigure}
	\caption{Event display for a $t\bar{t}$ event (a) and top-quark invariant mass computed from jets 
    reconstructed by FastJet or bSB (b). 
    Reproduced from Ref.~\citenum{qaiajet}. \href{https://creativecommons.org/licenses/by/4.0/}{CC BY 4.0}.
    \label{fig:mass} }
\end{figure}

\subsection{Summary and Outlook}

Foreseeing the HL-LHC to start operating in 2030 and other future 
colliders under consideration, innovations in computing are eagerly 
awaited. Along with the detector simulation, reconstruction, i.e., pattern recognition, 
is one of the most CPU-consuming tasks at high-energy colliders. 
Quantum artificial intelligence could potentially bring in significant speed-up
and/or allow us to perform previously unattainable computation.
Investigations have been continuously pursued in the past several years. 

With quantum circuit machines, various components of reconstruction could accelerate, 
with a caveat that quantum hardware needs to be improved in terms of error 
reduction, the presence of the QuAM architecture, and the number of qubits. The 
overhead of the classical data loading must also be carefully considered for 
practical applications~\cite{thaler1}. Quantum annealing allows us to handle 
larger datasets compared to the quantum circuit machines, but the current 
limited connectivity has been a crucial bottleneck for jet reconstruction. 
Meanwhile, quantum-inspired algorithms are promising for near-term applications, 
demonstrating their capability to handle large datasets and excellent 
computing speed and ground state predictions for 
Ising/QUBO problems. 

Nevertheless, both quantum circuit machines and quantum annealers have been continuously 
achieving important milestones in the past several years. Combined with 
High Performance Computing, quantum computing hardware may bring in paradigm shifts
in HEP in the coming years. However, some components of the experimental 
workflow require sub-million to million-level problem sizes, which are unlikely to 
be reached by error-tolerant quantum hardware in the near future. Quantum-inspired 
techniques are valuable regardless of the near and long-term success of the quantum 
hardware development. The rich program of quantum computing applications in 
HEP is likely to progress with constructive competitions among 
the three quantum technologies and other emerging techniques awaiting 
further development and practical implementations.

\section*{Acknowledgments}

HO is supported by 
the NSFC Basic Science Center Program for Joint Research on High Energy Frontier 
Particle Physics under Grant No.~12188102. 

\bibliographystyle{ws-mpla}
\bibliography{qai4hep}

\end{document}